\documentclass[conf]{new-aiaa}
\pdfoutput=1
\usepackage[utf8]{inputenc}
\usepackage{textcomp}
\usepackage{placeins}
\usepackage{graphicx}
\usepackage{amsmath}
\usepackage{multirow}
\usepackage[version=4]{mhchem}
\usepackage{siunitx}
\usepackage{longtable,tabularx}
\usepackage[acronym,nonumberlist]{glossaries}
\usepackage{hyperref}
\usepackage{csvsimple}
\usepackage{tabularx}
\usepackage{booktabs}
\usepackage{subcaption}
\usepackage{float}
\usepackage[table]{xcolor}
\hypersetup{
    colorlinks,
    citecolor=black,
    filecolor=black,
    linkcolor=black,
    urlcolor=black
}

\title{A Probabilistic Digital Twin of UK En Route Airspace for Training and Evaluating AI Agents for Air Traffic Control}

\author{Nick Pepper\footnote{Senior Research Associate, The Alan Turing Institute, London, England NW1 2DB, United Kingdom},  Adam Keane\footnote{Research Associate, The Alan Turing Institute, London, England NW1 2DB, United Kingdom}, Amy Hodgkin\footnote{Research Associate, The Alan Turing Institute, London, England NW1 2DB, United Kingdom},
Dewi Gould\footnote{Research Associate, The Alan Turing Institute, London, England NW1 2DB, United Kingdom}, Edward Henderson\footnote{Research Associate, The Alan Turing Institute, London, England NW1 2DB, United Kingdom},  \\Lynge Lauritsen\footnote{Research Software Engineer, The Alan Turing Institute, London, England NW1 2DB, United Kingdom} and Christos Vlahos\footnote{Research Software Engineer, The Alan Turing Institute, London, England NW1 2DB, United Kingdom}}  
\affil{The Alan Turing Institute, London, England NW1 2DB, United Kingdom}

\author{George {De~Ath}\footnote{Lecturer, Department of Computer Science, University of Exeter;} and Richard Everson \footnote{Professor, Department of Computer Science, University of Exeter;}}
\affil{University of Exeter}

\author{Richard Cannon\footnote{Senior Researcher, Department of Research and Development, NATS}, Alvaro Sierra Castro \footnote{Research and Development Graduate, Department of Research and Development, NATS}, John Korna\footnote{Researcher, Department of Research and Development, NATS}, Ben Carvell \footnote{Researcher, Department of Research and Development, NATS} and Marc Thomas \footnote{Researcher, Department of Research and Development, NATS; also Visiting Professor, Queen Mary University London}}
\affil{NATS, Fareham, England PO15 7FL, United Kingdom}

\begin{document}

\maketitle

\begin{abstract}
This paper presents the first probabilistic Digital Twin of operational en route airspace, developed for the London Area Control Centre. The Digital Twin is intended to support the development and rigorous human-in-the-loop evaluation of AI agents for Air Traffic Control (ATC), providing a virtual representation of real-world airspace that enables safe exploration of higher levels of ATC automation. 

This paper makes three significant contributions: firstly, we demonstrate how historical and live operational data may be combined with a probabilistic, physics-informed machine learning model of aircraft performance to reproduce real-world traffic scenarios, while accurately reflecting the level of uncertainty inherent in ATC. Secondly, we develop a structured assurance case, following the Trustworthy and Ethical Assurance framework, to provide quantitative evidence for the Digital Twin's accuracy and fidelity. This is crucial to building trust in this novel technology within this safety-critical domain. Thirdly, we describe how the Digital Twin forms a unified environment for agent testing and evaluation. This includes fast-time execution (up to x200 real-time), a standardised Python-based ``gym'' interface that supports a range of AI agent designs, and a suite of quantitative metrics for assessing performance. Crucially, the framework facilitates competency-based assessment of AI agents by qualified Air Traffic Control Officers through a Human Machine Interface. We also outline further applications and future extensions of the Digital Twin architecture.

\end{abstract}
\newpage
%\section*{Glossary}
%{\renewcommand\arraystretch{1.0}
%\noindent\begin{longtable*}{@{}l @{\quad=\quad} l@{}}
%AIRAC & Aeronautical Information Regulation and Control \\
%ATCO & Air Traffic Control Officer \\
%ATC  & Air Traffic Control \\
%AIP & Aeronautical Information Publication \\
%ATM  & Air Traffic Management \\
%API & Application Programming Interface\\
%BADA & Base of Aircraft Data \\
%CAA & Civil Aviation Authority (UK) \\
%CAS & Calibrated Airspeed \\
%DT & Digital Twin \\
%EASA & European Union Aviation Safety Agency \\
%ECMWF & European Centre for Medium-Range Weather Forecasts \\
%FAA & Federal Aviation Authority (US) \\
%FL & Flight Level \\
%GSN & Goal Structured Notation \\
%HMI & Human Machine Interface \\
%IAS & Indicated Airspeed \\
%ISA & International Standard Atmosphere \\
%LAC & London Area Control \\
%KS & Kolmogorov-Smirnov \\
%MAE & Mean Absolute Error \\
%MET & Meteorological \\
%ML & Machine Learning \\
%NMI & Nautical Mile \\
%PIML & Physics-informed Machine Learning \\
%PSR & Primary Surveillance Radar \\
%ROCD & Rate of Climb/descent \\
%SSR & Secondary Surveillance Radar \\
%TAS & True Airspeed \\
%TEA & Trustworthy and Ethical Assurance \\
%TP & Trajectory Prediction \\
%WAM & Wide Area Multilateration \\
%\end{longtable*}}

\section{Introduction}

\lettrine{G}rowing demand for air travel, with flight numbers rising steadily each year~\citep{easr_2022}, is placing significant pressure on Air Traffic Management (ATM) systems to increase capacity while maintaining rigorous safety standards. This challenge is intensified by the introduction of new airspace users such as drones and electric Vertical Take-Off and Landing (eVTOL) aircraft, as well as by governmental requirements to improve airspace efficiency and reduce emissions (see, e.g.,~\citep{jetzero}). In response, major ATM modernisation programmes~\citep{faa, sesar} aim to improve workload forecasting, trajectory prediction accuracy, strategic planning, route optimisation, and overall system resilience~\citep{Sesar_sol}. The scale and urgency of these challenges mean that incremental enhancements to traditional tools and processes are unlikely to be sufficient, and a step change in automation for tactical air traffic control will be required. 

Within this context, there is growing interest in AI-based decision-support technologies for Air Traffic Control (ATC). Throughout this paper, we use the term ``AI agent'' to denote any AI-based decision-support technology for tactical ATC, irrespective of its envisioned level of autonomy. Such agents may range from tools that highlight potential conflicts and propose resolutions to more advanced systems with greater degrees of control authority~\citep{argos, easa_guidelines_2024}. ATC is a safety-critical task, requiring controllers to balance efficiency, orderliness, and other factors in a highly context-specific manner, while also maintaining rigorous safety standards. In consequence, AI agents require extensive and rigorous validation before operational deployment. Automated methods alone, whether formal methods~\citep{munoz2015formal} or the recent \emph{validation-through-scenario-generation} paradigm that has emerged in autonomous driving~\citep{AV_scengen_review, self_driving_cars} are insufficient, as they cannot capture the situational judgements made by trained controllers. Human-in-the-Loop (HITL) assessment~\cite{Agent_validation_framework} is therefore essential, enabling qualified controllers to interrogate agent behaviour in realistic conditions and against appropriate competency frameworks.

HITL evaluation must faithfully represent the uncertainty inherent in ATM operations. At a system level, uncertainty arises from schedule deviations~\citep{pyrgiotis2013modelling, sridhar2009modeling, route_uncertainty1, route_uncertainty2}, convective weather~\citep{michalek2009identification, matthews2010assessment}, and unexpected events such as loss of communication, military activity, or hijacking~\citep{atc_emergencies}. At the aircraft level, uncertainty in mass~\citep{sun2018aircraft}, operator procedures~\citep{amy_aiaa}, pilot intent, and local atmospheric conditions~\citep{met_sensitivity_tp} contributes to substantial variability between predicted and observed trajectories~\citep{pepper4984556probabilistic, aerospace9020091}. Deterministic simulations therefore provide an incomplete basis for evaluating AI agent behaviour under realistic operational conditions, particularly when the agents themselves must reason under uncertainty.

A simulation environment suitable for AI agent development and HITL assessment must therefore satisfy several requirements:
\begin{itemize}
    \item \textbf{A high-fidelity digital representation of airspace} that captures sectorisation, route structure, procedures, and other factors that form the operational context for an air traffic sample.
    \item The capacity for \textbf{probabilistic simulation of aircraft performance} and traffic evolution, in order to represent both aleatoric variability and the level of epistemic uncertainty inherent in ATM operations.
    \item An \textbf{interface for AI agents to perform tactical control under uncertainty}, enabling them to perceive, decide, and act within the simulated environment.
    \item \textbf{Quantitative metrics} to assess AI agents, together with a \textbf{Human Machine Interface (HMI)} that enables competency-based assessment by human instructors.
    \item An \textbf{interface with the live or recorded real-world operational environment}, allowing AI agent behaviours to be assessed on real-world traffic samples and scenarios.
    \item The facility to \textbf{run significantly faster than real-time} as an enabler for AI agents that employ Machine Learning methods such as Reinforcement Learning. These techniques require many thousands of hours of simulation in order to train effective policies.
\end{itemize}

Existing ATC simulators and open source systems offer valuable capabilities, but do not simultaneously provide probabilistic aircraft performance modelling, integration with operational data streams, and interfaces designed for AI agent control and HITL assessment~\citep{hoekstra2016bluesky, loft2004atc, bilimoria2001facet, eurocontrol_escape, eurocontrol_edep, molina2014agent, vatsim, atc_sim, openScope}. These limitations motivate the development of a probabilistic Digital Twin.

Digital Twins provide high-fidelity, data-driven virtual representations of physical systems and have seen growing adoption in safety-critical domains such as healthcare~\citep{katsoulakis2024digital}, energy~\citep{yu2022energy} and transportation~\citep{wu2022digital, kuvsic2023digital, kim_aerodts}. Probabilistic Digital Twins, which explicitly model the uncertainties of the system, have been explored in aerospace for structural and component modelling including wings~\citep{li2017dynamic}, unmanned aerial vehicles~\citep{kapteyn2021probabilistic}, and rotorcraft systems~\citep{prob_rotorcraft_DT}. At the airspace level, \citet{drones7070484} consider a Digital Twin for future airspace concepts focused on uncrewed and urban air mobility operations below 400~ft. To the authors’ knowledge, however, there is currently no probabilistic Digital Twin of an operational en route airspace designed specifically to support AI agent development, benchmarking, and HITL evaluation. This work makes the following contributions:
\newpage

\begin{itemize}
    \item \textbf{A probabilistic Digital Twin of operational en route airspace}, integrating multiple operational data sources with a probabilistic, physics-informed aircraft performance model.
    \item \textbf{A quantitative validation of the Digital Twin}, demonstrating trajectory accuracy, uncertainty fidelity, and replication of human ATCO training exercises.
    \item \textbf{A unified framework for AI agent development, training and evaluation}, including automated metrics and interfaces for AI agent control of aircraft and HITL assessments.
\end{itemize}

This paper introduces a probabilistic Digital Twin for the London Area Control Centre (LACC) environment. The Digital Twin integrates real-world surveillance, flight plans, sectorisation, coordination procedures, and meteorological data with a probabilistic, physics-informed machine learning (PIML) model of aircraft performance to reproduce realistic traffic patterns and operational uncertainty. The environment supports AI agent development and evaluation under realistic conditions and enables quantitative and competency-based assessment through HITL exercises. The remainder of this paper is structured as follows: Section~\ref{sec:methods} describes the core components of the probabilistic Digital Twin. Section~\ref{sec:assurance} outlines an assurance framework regarding the suitability of the twin as an agent training and evaluation platform and presents quantitative results regarding its accuracy and fidelity. Section~\ref{sec:future} describes other use cases and future plans for the Digital Twin, and Section~\ref{sec:conclusion} concludes.

\section{Methodology}\label{sec:methods}

This section describes how the three main contributions outlined in the introduction are realised in practice. Firstly, we detail the construction of a probabilistic Digital Twin of the LACC environment that integrates multiple operational data sources with a probabilistic, PIML aircraft performance model. Secondly, we describe the unified framework that this twin provides for AI agent development, training and evaluation, including interfaces for both automated agents and human-in-the-loop assessment. The third contribution, a quantitative validation of the Digital Twin, is addressed in Section~\ref{sec:assurance}, where we present evidence regarding its accuracy and fidelity within an assurance framework.

\begin{figure}[ht]
    \centering
    \begin{minipage}{1.0\linewidth}
        \centering
    \includegraphics[width=1.0\textwidth]{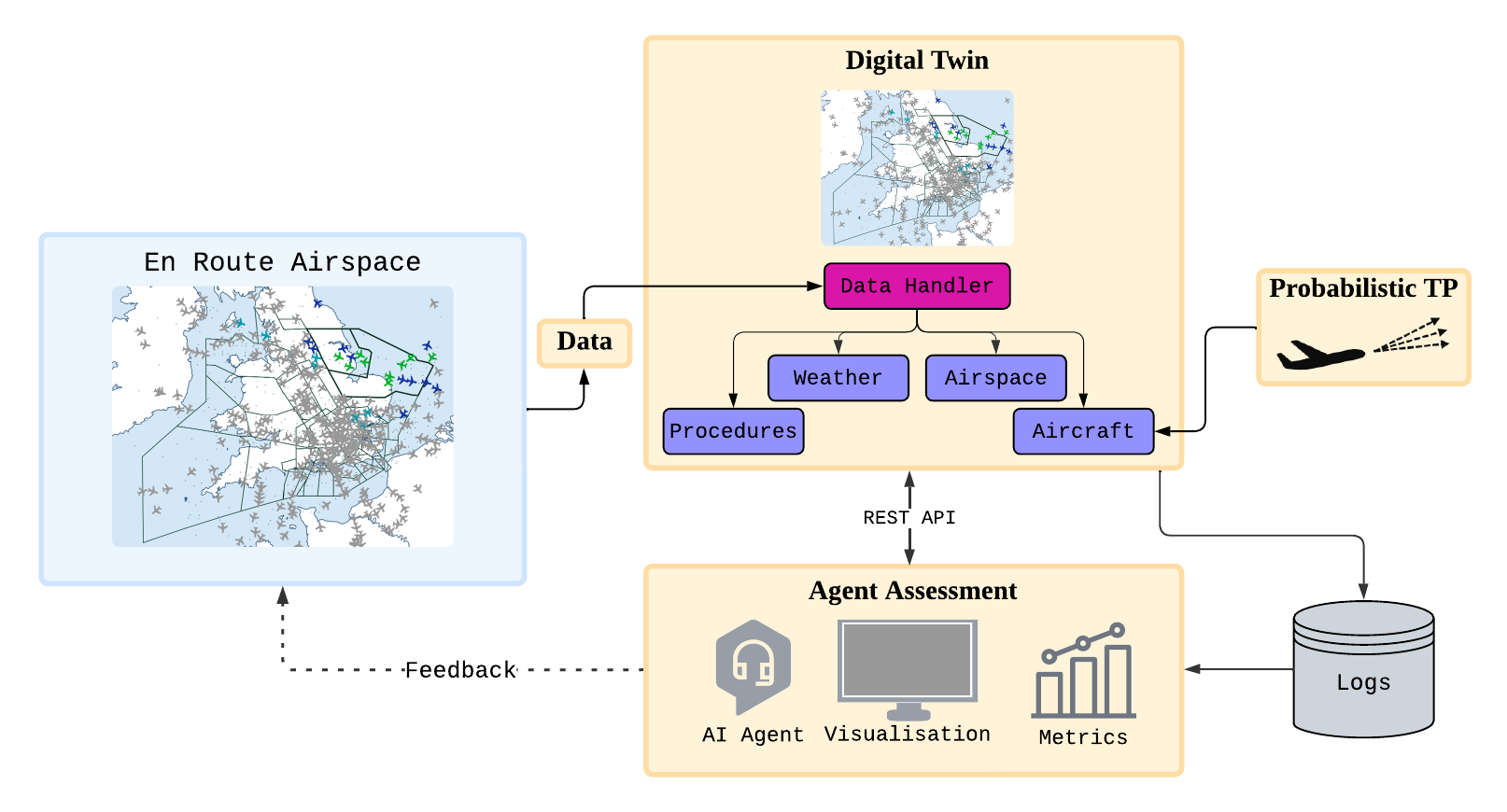}
        \caption{A schematic outlining the main components of the Digital Twin, the probabilistic trajectory predictor, and agent assessment framework.}
        \label{fig:DT_schematic}
    \end{minipage}
\end{figure}

The Digital Twin reproduces operational configurations, procedures, and traffic patterns by integrating historical and live operational data from NATS' systems. Aircraft motion is modelled either from recorded flight trajectories or through a probabilistic trajectory prediction (TP) engine that produces realistic aircraft behaviour. This approach, combining observed and simulated trajectories, enables controlled experimentation whilst reproducing real-world traffic patterns and flight dynamics. The system includes an interactive HMI designed to closely resemble current controller displays, and a suite of automatically generated performance metrics to capture safety, efficiency, and conformance to operational procedures. Together, these support a systematic assessment of agent decision-making. Figure~\ref{fig:DT_schematic} is a schematic that outlines the various components of the Digital Twin system. Key components are described in greater detail in the following subsections.

\subsection{Data Sources}
The Digital Twin integrates surveillance data, flight plans, controller clearances, aircraft communications status, airspace configurations, and meteorological information from NATS' operational systems. A comprehensive dataset of over 20 million flights is available, providing high-fidelity, near-continuous historical coverage from 2016 onwards. In addition, a live data feed enables the twin to replicate current traffic conditions with minimal latency.

Aircraft surveillance data is provided by ARTAS~\citep{eurocontrol_artas_web}, a data fusion and tracking system combining inputs from PSR (Primary Surveillance Radar), SSR Mode A/C/S (Secondary Surveillance Radar), ADS-B and Wide Area Multilateration (WAM). This supplies aircraft position, altitude, heading, and velocity at approximately 6 second intervals, along with additional information when available, such as the selected Flight Level (FL). Flight plans, controller-issued clearances, sector configuration information, and sector coordination records are sourced from NATS' NERC en route air traffic management system which provides timestamped event-level updates.

Historical and current airspace structure and waypoint information are drawn from the Aeronautical Information Publication (AIP) and NATS' internal systems, allowing an accurate, date-dependent definition of the operational environment in which aircraft and controllers interact. In order to replicate the meteorological (MET) information available to controllers in the live operational system, operational forecasts, originating from the UK Met Office at three-hour intervals, are made available within the Digital Twin framework. Furthermore, to enable accurate modelling of aircraft performance and wind-relative motion in historic scenarios, the Digital Twin uses European Centre for Medium-Range Weather Forecasts (ECMWF) Reanalysis v5 (ERA5)~\citep{ERA5} MET datasets, which provides hourly high-resolution estimates of weather and climate by combining observations with numerical prediction models.

\subsection{Digital Twin}\label{subsec:fidelity}

The Digital Twin provides a digital representation of the en route controlled airspace of LACC. Its purpose is to recreate the operational environment with sufficient realism to support the development, testing, and evaluation of AI agents under realistic traffic patterns and constraints. Fidelity is achieved through Aeronautical Information Regulation and Control (AIRAC) aligned airspace definitions, data-driven trajectory replay and aircraft presentation to sectors, realistic controlling procedures and a probabilistic, PIML model of aircraft dynamics.

The airspace is partitioned into interlocking polyhedral regions, known as sectors~\citep{pepper4984556probabilistic}. These form the fundamental units of controller responsibility. In daily operations, the 32 base sectors may be dynamically combined into larger \textit{bandboxed} configurations to balance workload. Each active bandboxed sector group is normally managed by a single tactical controller, often supported by a planning controller. Depending on traffic demand, between 5 and 15 bandboxed sectors are typically active at any time. The airspace representation includes all UK airports and relevant navigation points (waypoints, fixes, and navaids). These elements define planned and executed routes in filed flight plans, as well as procedural constraints (e.g., requiring a descent to a defined altitude by a designated point). For historical scenarios, sectors within the airspace are date aligned to their AIRAC definitions.

Each aircraft is represented as an object with physical state variables (altitude, position, heading, and airspeed), and operational attributes (callsign, active clearances, intent, and controlling sector). Flight plan information, describing the aircraft's type, requested flight levels and speeds, departure and destination aerodromes, and planned route is also included. Controllers (AI or human) interact with aircraft through the set of clearances in Table \ref{tab:clearance}, which reflect those commonly used in LACC. A simple pilot agent mediates between issued clearances and the aircraft's intent state. This allows realistic instruction delays, and can also be parametrised to create extreme or unusual events such as clearance modification due to misinterpretation.

\begin{table}[h]
    \centering
    \caption{Action space of the Digital Twin.}
    \begin{tabularx}{\textwidth}{ p{4cm}| p{4cm} |X }
        \toprule
        Clearance & Attributes & Description \\
        \midrule
        Route direct to & Waypoint & Fly direct to a specified waypoint and continue route from that point \\
        Fly heading & Heading & Fly the specified heading\\
        Turn left/right by degrees & $\Delta$Heading & Turn left or right by the specified amount in degrees \\
        Maintain present heading & - & Fly the current aircraft heading \\
        Climb/descend now & Flight level & Climb or descend immediately to the specified flight level \\
        Descend when ready to be level by a waypoint & Flight level and waypoint & Descend at optimal top-of-descent such that the given level is achieved at or abeam the specified waypoint \\
        Descend now to be level by a waypoint & Flight level and waypoint & Descend immediately to achieve the given level at or abeam the specified waypoint \\
        Change CAS to & CAS (kt) & Changes the calibrated airspeed used below the speed transition altitude \\
        Change Mach speed to & Mach speed & Changes the  Mach speed used above the speed transition altitude \\
        Change rate of climb/descent to & Vertical speed (ft/min) & For an aircraft already climbing or descending, modifies its rate of climb/descent \\
        Contact another frequency & Next frequency or sector & Transfer to the specified new sector frequency\\
        \bottomrule
    \end{tabularx}
    \label{tab:clearance}
\end{table}

Sector to sector handovers are governed by \textit{coordinations}, which specify agreed transfer levels, points, and estimated handover times. Coordinations may be predefined through automated or route-dependent rules, or can be negotiated tactically. The Digital Twin represents these agreements and allows them to be modified. The transfer of communications for an aircraft between sectors is triggered by instructions that emulate real-world frequency changes, either through data-driven events or manual instructions.

Wind vectors are modelled on a configurable three-dimensional grid. Agents access forecast wind fields through an API, while the ``true'' wind used in aircraft dynamics remains hidden. Both forecast and actual winds may be sourced from data or artificially generated.

The Digital Twin is designed to establish a safe, high-fidelity environment dedicated to the development and rigorous testing of AI agents, particularly when subject to realistic traffic patterns and operational constraints. The system achieves this by integrating data replay and simulation within a single framework, thereby enabling real-world boundary conditions to be replicated while allowing full AI agent control inside selected sectors within the airspace. At any time, a user may define one or more \textit{simulated sectors}, within which aircraft can be controlled, and their dynamics are generated by the twin. Aircraft outside these sectors are replayed directly from historical or live data, with their positions and operational attributes updated to match real-world trajectories. When an aircraft's control passes to a simulated sector, its dynamics transition to a probabilistic simulator (Section~\ref{subsec:prob_TP}), and remain simulated until it exits the airspace.

\begin{figure}[ht]
    \centering
    \begin{minipage}{0.8\linewidth}
        \centering
        \includegraphics[width=\linewidth]{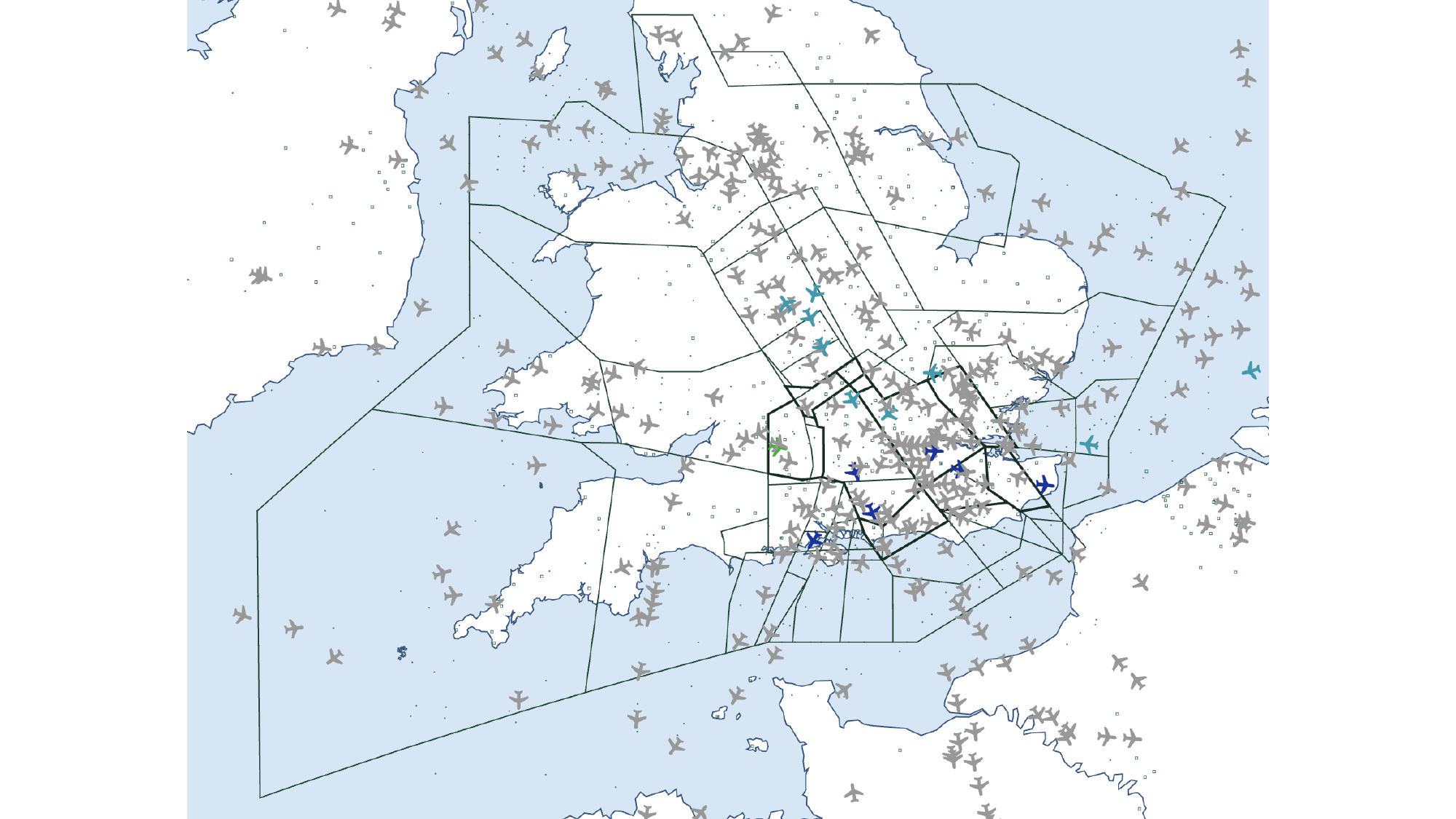}
        \caption{London Area Control Sectors and traffic sample replayed in the Digital Twin.}
        \label{fig:LondonAreaControl}
    \end{minipage}
\end{figure}

The Digital Twin also supports simultaneous human control of adjacent sectors, enabling studies in human-machine teaming. Comprehensive logging captures all system states, facilitating full scenario playback and detailed post-scenario analysis. The system architecture that underpins these capabilities is shown in Figure~\ref{fig:DT_schematic}. Additional modes of operation, such as fully data driven, fully simulated, or hybrid (e.g., simulated trajectories with data-driven clearances), extend the application of the twin beyond the use cases examined in this paper and are briefly discussed in Section~\ref{sec:future}. Figure~\ref{fig:LondonAreaControl} shows a historic traffic sample replayed in the Digital Twin, illustrating aircraft positions and the 32 LACC sectors.

\subsection{Probabilistic Aircraft Trajectories}\label{subsec:prob_TP}

The primary purpose of the Digital Twin is to enable the development and evaluation of AI agents under conditions that faithfully reflect the real-world. In operations, aircraft trajectories are influenced by aleatoric and epistemic sources of uncertainty, including localised weather, variations in aircraft mass and operating procedures, and differences in pilot intent and timing. When inferred solely from surveillance and air traffic control data, these effects render future aircraft trajectories inherently unpredictable.

For agent development, reproducing this uncertain behaviour within the Digital Twin is essential for two closely related reasons linked to generalisability. Firstly, deterministic simulations cause many machine learning methods to implicitly learn the precise, noise-free trajectories observed during training, which can lead to poor generalisation when agents are exposed to real traffic. Secondly, because air traffic control is a safety-critical domain, any assessment of an agent's safety performance must account for the full range of realistic aircraft performance, including improbable but plausible climb rates, speeds, and response delays, not only nominal conditions. A comprehensive safety evaluation would also consider rare or abnormal events such as radio failures or airspace incursions. These are discussed briefly in Section~\ref{sec:future} on future work. 

In order to model the effects of uncertainties, the Digital Twin generates trajectories using a probabilistic PIML model trained on historic data. The Base of Aircraft Data (BADA) model~\citep{nuic2010user}, a performance model developed and maintained by EUROCONTROL, provides altitude-dependent parameters for each aircraft type and a set of energy-balance equations that are solved numerically to generate deterministic climb and descent profiles. Our approach extends BADA by replacing its Calibrated Airspeed (CAS), thrust, and drag parameters, which are functions of the flight level, with a generative model constructed using functional principal component analysis and Gaussian mixture models trained on historic trajectories. This physics-informed method allows the Digital Twin to produce plausible trajectories probabilistically, with the BADA model's energy constraints ensuring physical plausibility. Fitting the generative components to local historical data also tunes the model to the characteristics of a specific airspace, resulting in generated trajectories that more closely resemble observed ones. A full technical description is provided in \citet{Pepper_DataBADA} and \citet{hodgkin2025probabilisticsimulationaircraftdescent}. Cruise speeds are generated probabilistically by replacing BADA-defined cruise CAS and Mach with values sampled from the probability mass function of historic data.

\subsection{AI Agent Assessment Framework}\label{subsec:agent_assessment}

This section describes the assessment framework developed to support the training, benchmarking, and evaluation of AI agents for en route ATC. The framework extends the baseline Digital Twin environment with fast-time execution capabilities, quantitative performance metrics, a human-in-the-loop (HITL) evaluation interface, and scalable scenario generation tools. These components have collectively enabled systematic and reproducible analysis of diverse agent designs, including rules-based approaches~\citep{hawk,mallard}, Reinforcement Learning~\citep{RL_paper} and optimisation-based methods (see~\citet{Agent_validation_framework} for further detail).

\textbf{Fast-Time Execution for Agent Training}: A key requirement for interaction-intensive agent training, particularly Reinforcement Learning that learns through trial-and-error, is the ability to run simulations in \textit{fast-time} mode. The Digital Twin achieves approximately $\times 10$ real-time speed for full LACC simulations, $\times 50$ for individual sectors, and up to $\times 200$ for the simplest scenarios, depending on traffic density. Fast-time execution preserves the same dynamic and procedural fidelity as real-time mode, enabling large-scale exploration without compromising operational consistency.
\newpage
\textbf{Quantitative Metrics for Automated Evaluation}: Agent performance is assessed using quantitative metrics computed at runtime to support automated evaluation. These include \textbf{\textit{safety metrics}} (loss of separation counts, assured-safety indicators), \textbf{\textit{efficiency metrics}} (fuel consumption rate, the UK 3D Inefficiency Score (3Di)~\citep{nats2012fuel}), and procedural metrics (compliance with coordination conditions, number of clearances issued). Metrics are also included that support the development of specific classes of AI agent. For instance, a series of objectives is provided for Reinforcement Learning agents that combine domain-driven performance measures with engineered gradients and robustness to exploitation. While these metrics provide objective measures of tactical performance, they do not fully capture the richness of operational decision-making or the appropriateness of agent intent. Accordingly, they are complemented by a dedicated HITL assessment framework.

\begin{figure}
    \centering
    \includegraphics[width=1.0\linewidth]{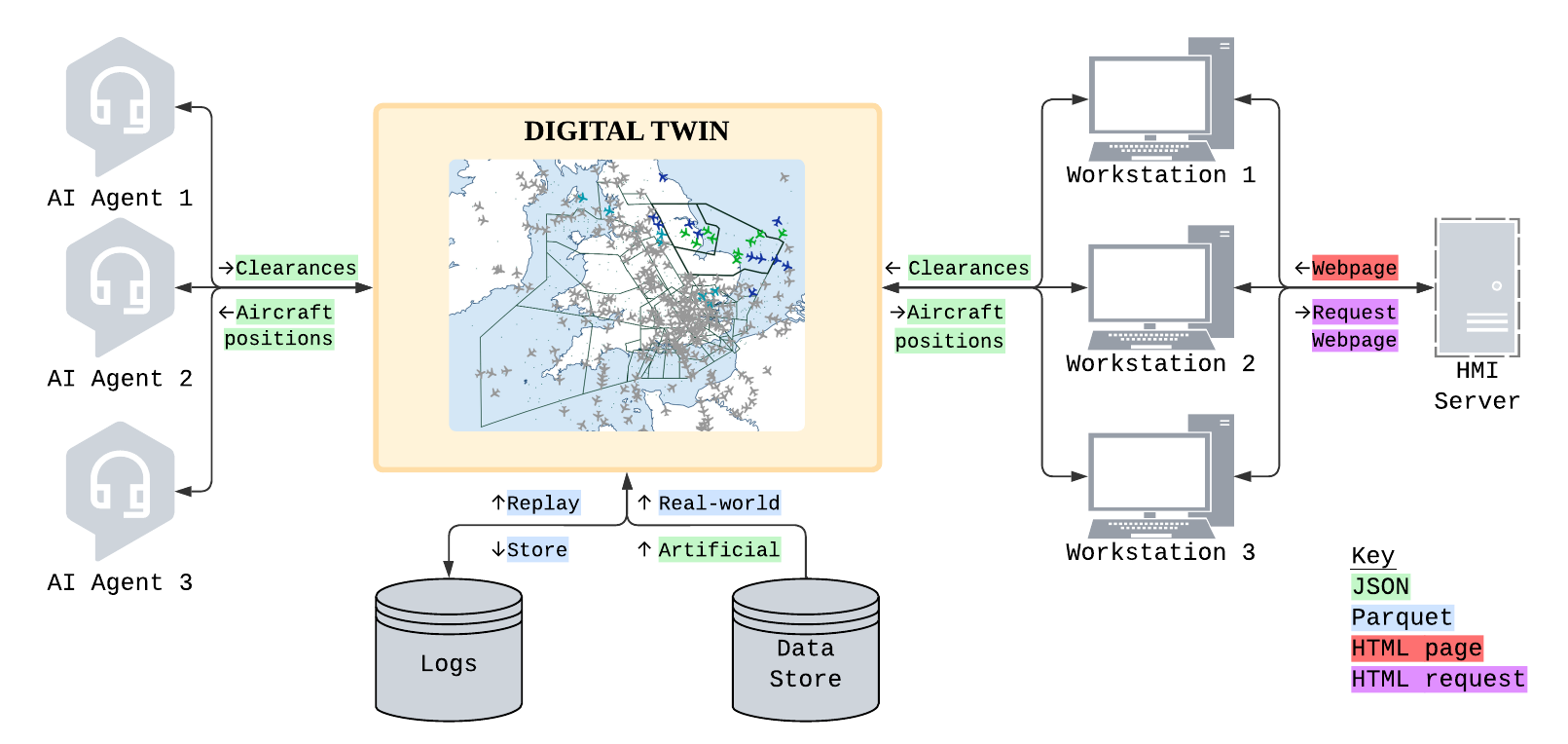}
    \caption{Diagram of API Framework for Agent Assessment.}
    \label{fig:API_Framework}
\end{figure}   

\textbf{Human-in-the-Loop (HITL) Operational Assessment}: This HITL framework enables qualified ATCOs to observe, interrogate, intervene in, and replay agent-driven simulations. It provides real-time visualisation tools, including an operational-style radar display, digital flight strips, and an action panel modelled on those used in live operations. An underlying server-based architecture allows ATCOs and AI agents to interact concurrently with the same simulation instance. Controllers may assume control of individual aircraft, inspect agent intent through exposed planned manoeuvres and trajectories (when supplied by the agent), and offer qualitative feedback on agent decision-making. A schematic overview is provided in Figure~\ref{fig:API_Framework}.

\textbf{Scenario Generation for Scalable Experimentation}: To enable controlled and repeatable experimentation, a \textit{scenario generation} module was implemented. This tool allows the creation of arbitrarily many parametrised scenarios derived either from real-world sector data or from simplified artificial geometries. Parameters include traffic density, conflict geometry, weather conditions, and boundary configurations. All scenarios are reproducible through seeded randomisation.

\textbf{Agent ``Gym'' Interface for Training and Benchmarking}: To simplify agent interaction with the Digital Twin, a dedicated Python-based agent gym interface~\citep{openai_gym} was developed, supporting Reinforcement Learning as well as rules-based, optimisation-based, and hybrid agent designs. The module provides a formalised API for stepwise environment interaction, defining permissible actions, abstracting simulation state, and exposing a common set of reward signals and runtime metrics. By decoupling agent logic from simulator internals and enforcing a consistent interface, the gym lowers the barrier for agent development, enables rapid prototyping, and supports scalable benchmarking across heterogeneous agent types. This unified interface also facilitates controlled training pipelines, reproducible testing regimes, and systematic comparison of agent performance across a range of scenarios.

\textbf{Structured Assessments via NATS' ATCO Assessment Course Replication}: In addition to data-driven and user-defined scenarios, a structured agent assessment environment was created by reproducing NATS' internal (human) trainee ATCO assessment course within the Digital Twin. This environment comprised an artificial sector and hand-crafted formative and summative scenarios developed by NATS training specialists to evaluate core tactical control competencies. A trajectory generation subsystem replicates the aircraft dynamics of NATS' in-house simulator to ensure that each scenario presents the intended operational challenges and difficulty profile. In human training, successful completion of this course, typically after five months of full-time instruction, is required for progression to advanced training. For AI research, this regulator-certified scenario set provides a controlled and pedagogically curated benchmark for analysing agent strengths, weaknesses, and failure modes, complementing broader testing on real-world traffic data. A detailed discussion of this agent assessment framework is provided in \citet{Agent_validation_framework}.

\section{Assuring the Accuracy and Fidelity of the Digital Twin}\label{sec:assurance}

Digital Twins pose substantial challenges to existing assurance pathways in ATM. They integrate simulation and bidirectional data exchange with operational systems, and may include interoperation with decision support or control agents. Although regulatory guidance exists for certain underlying technologies, such as the use of simulation within synthetic training devices~\citep{easa_atco_training}, there is currently no comprehensive framework that directly addresses the assurance of Digital Twins in ATM. This gap is particularly significant when Digital Twins are capable of influencing real-world operations through the decision-making of AI agents. In this case, it is unclear whether the Digital Twins and AI agents should be assured separately or jointly, given the potential emergent behaviour arising from their interaction.

The assurance problem is further complicated by the use of machine-learned components within Digital Twins, such as the probabilistic trajectory generator described in Section~\ref{subsec:prob_TP}. Although regulators have begun publishing high-level principles for the use of ML in ATM workflows~\citep{caa_cap2970, faa_roadmap}, no detailed published standards exist for the use of either ML models or Digital Twins in this domain. Existing academic work on Digital Twin validation predominantly proposes generic, domain-agnostic frameworks~\citep{dt_validation, wagg2020digital, nasem2024}, while existing domain-specific standards remain focused on deterministic methods. As a result, current guidance is insufficient to support robust assurance activities for probabilistic, ML-enabled Digital Twins such as the one presented here.

To address these shortcomings, the Trustworthy and Ethical Assurance (TEA) platform has been proposed to construct structured, argument-based assurance cases for Digital Twins~\citep{burr2024}. The platform employs Goal Structured Notation (GSN), allowing users to define top-level Goal Claims and decompose them into Strategies. These strategies are supported by Property Claims, contextual information, and Evidence elements, which may be either quantitative or qualitative. Evidence and Property Claims are themselves constructed from Assumption and Justification elements about their implementation. Notably, an assurance case addressing the accuracy and fidelity of the Digital Twin discussed in this work has been developed by \citet{adam_validation}. 

In this project, the Digital Twin functions as a training and development platform for AI agents. Consequently, the assurance case in \citet{adam_validation} focuses on establishing the accuracy and fidelity of the Digital Twin itself, which is a necessary prerequisite for the safety-assurance activities required for operational deployment. The Digital Twin is therefore evaluated here as a stand-alone system, serving as a precursor to future assurance work regarding its interaction with AI agents. Rather than reproducing the full assurance case, this paper presents quantitative evidence supporting two central Property Claims: (i) the accuracy of the probabilistic simulation of real-world sectors, and (ii) the fidelity with which formative exercises from NATS' ATCO assessment course have been replicated. Both claims are fundamental to validating the suitability of the Digital Twin as a training and testing environment.

\subsection{Accuracy of the probabilistic simulation of real-world sectors}\label{sec:assure_prob}

\begin{figure}
    \centering
    \includegraphics[width=0.8\linewidth]{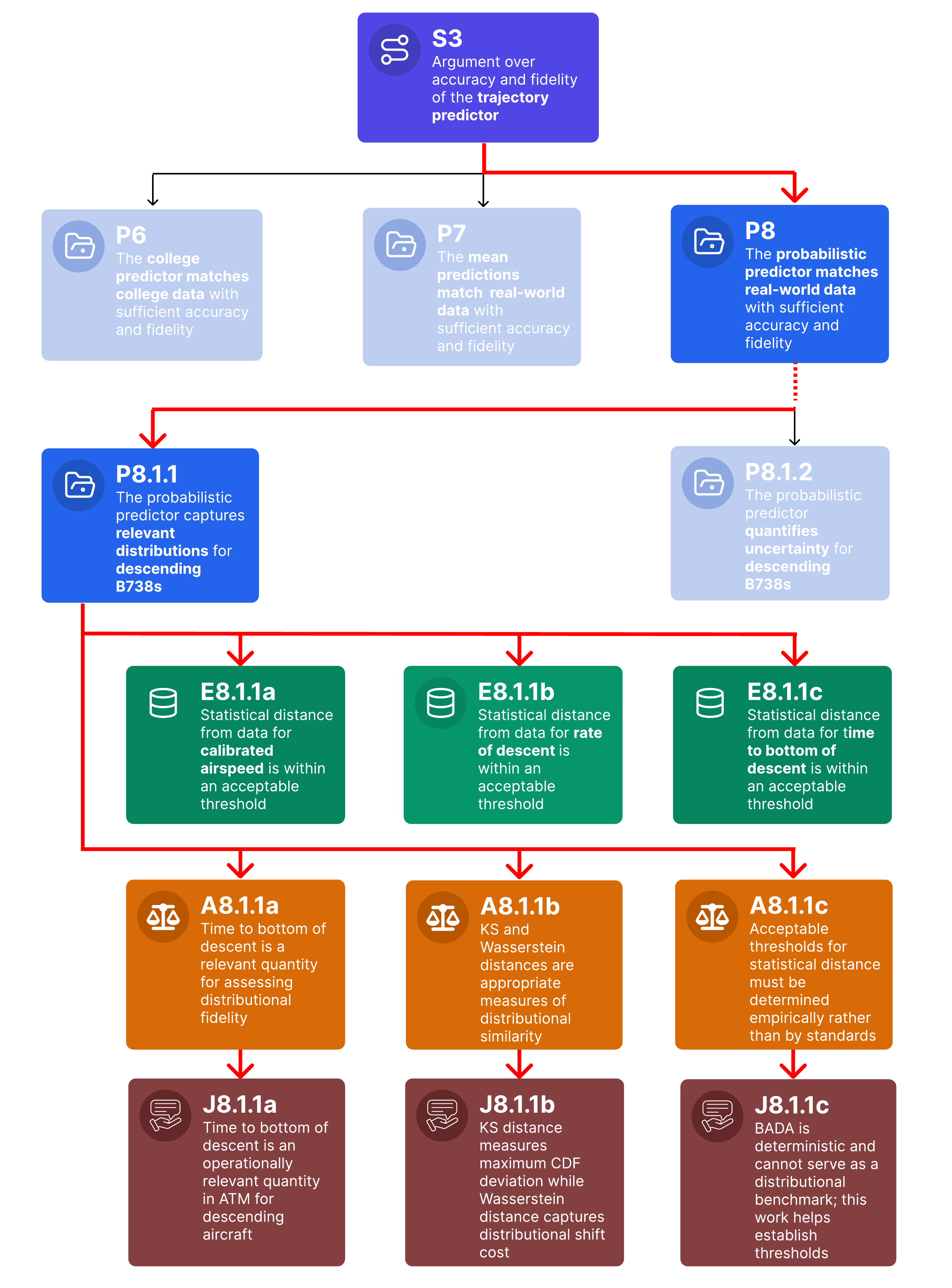}
    \caption{A subset of the GSN tree built for the assurance case presented in \citet{adam_validation}. It displays those strategies and highlights some property claims that are relevant to the task of assuring the probabilistic trajectory generator.}
    \label{fig:assurance_real_world}
\end{figure}

In this section, we discuss some key Property Claims developed in \citet{adam_validation} relating to the assurance of the probabilistic trajectory predictor in the Digital Twin. Figure~\ref{fig:assurance_real_world} shows a subset of the GSN, beginning with a top-level Strategy, which groups related Property Claims, in this case, relating to the accuracy and fidelity of the twin. The P8 Property Claim relating to the accuracy of the probabilistic trajectory predictor is highlighted\footnote{ Each Property Claim has a unique numerical ID; lower-numbered Property Claims relate to other strategies S1 and S2 e.g.\ around the data pipeline. See \citet{adam_validation} for details.}. The full tree under P8 is extensive, with sub-Property Claims, Evidence, Assumptions, and Justifications for every commonly occurring aircraft type and phase of flight beneath this top-level property claim. For readability, the tree is trimmed, with only sub-Property Claims relating to the Boeing 737-800 (B738) being displayed. Similarly, Evidence, Assumptions, and Justifications are displayed for only the B738 in the descent sub-Property Claim. The following sections present evidence that could be used to support the sub-Property Claims in Figure~\ref{fig:assurance_real_world}. 

\subsection*{Capturing real-world uncertainty}
Property Claim 8.1.1 relates to how well sampled trajectories drawn from the probabilistic trajectory predictor capture the variability in real-world trajectories for descending B738s. Evidence is provided for this Property Claim by sampling trajectories from the probabilistic trajectory predictor and comparing them in distribution with a test dataset of held-out trajectories. 

\begin{figure}
    \centering
    \includegraphics[width=0.8\linewidth]{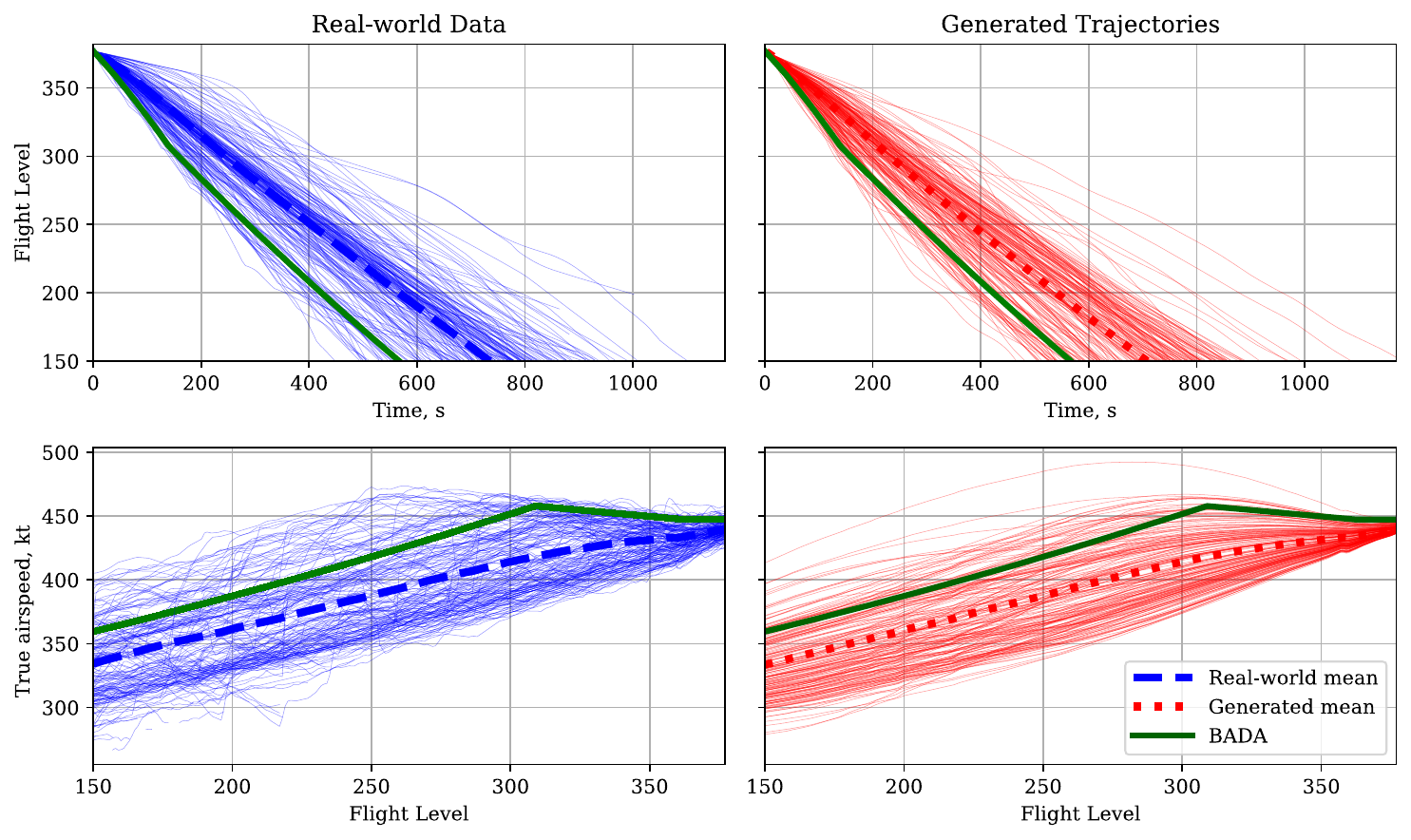}
    \caption{A held out test dataset of B738 data (left) versus generated trajectories using the probabilistic trajectory predictor (right).}
    \label{fig:gen_descent_B738}
\end{figure}

Figure~\ref{fig:gen_descent_B738} displays generated descending B738 trajectories, with the right-hand panels displaying trajectories that are generated by the probabilistic trajectory generator. The top-right panel shows FL against time and the bottom-right True Airspeed (TAS) against FL for the generated trajectories. In contrast, the held out test dataset of B738 data is displayed in the left panels. This aircraft type and phase of flight was chosen as an example because the B738 is the most common aircraft in LACC and descents are the most challenging phase of flight to model accurately. This is due to the variations in aircraft performance that arise primarily from airspace and airline procedures~\citep{hodgkin2025probabilisticsimulationaircraftdescent, pepper4984556probabilistic}. From visual inspection, there is good agreement between the model means and those of the data, particularly when compared to the BADA-generated trajectory (green), which can be significantly misspecified in descent for the reasons described above. The set of descending trajectories appears to match well in the upper plots displaying FL against time. The TAS of the generated trajectories also appear to match well in overall distribution, although the line representing each trajectory tends to be smoothed due to the functional principal component analysis that is performed to reduce the dimensionality of the data. 

\begin{figure}
    \centering
    \begin{subfigure}{0.5\textwidth}
    \includegraphics[width=1\textwidth]{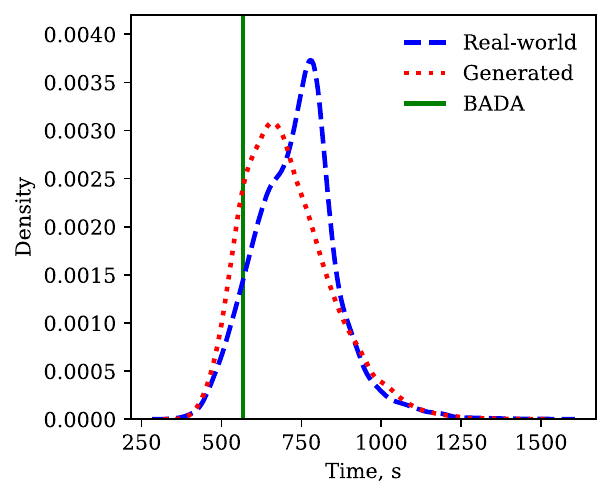}
    \caption{PDF}
    \end{subfigure}%
    \begin{subfigure}{0.468\textwidth}
    \includegraphics[width=1\textwidth]{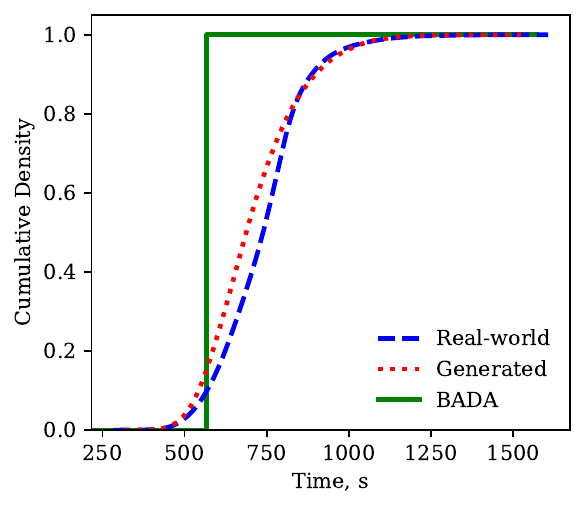}
    \caption{CDF}
    \end{subfigure}
    \caption{Distribution of time to bottom of descent for real-world and generated trajectories for the B738.}
    \label{fig:B738_time_to_descend}
\end{figure}

Evidence for the accuracy of the probabilistic trajectory predictor may then be generated by computing the statistical distance between the distributions of the two sets of trajectories. In \citet{Pepper_DataBADA} and \citet{hodgkin2025probabilisticsimulationaircraftdescent} the accuracy of the probabilistic trajectory generator in climb and descent, respectively, is benchmarked for quantities such as the top of climb (or bottom of descent) and the ROCD and CAS at specified flight levels. Probability distributions for these quantities are inferred from the set of trajectories. As an example of this evidence, the left panel of Figure~\ref{fig:B738_time_to_descend} displays probability density functions for the time to bottom of descent for the B738, while the right-hand panel displays the corresponding cumulative distribution functions. Visually, there is good agreement between the two distributions, which can be quantified using metrics such as the Kolmogorov-Smirnov (KS) distance and the Wasserstein distance (A8.1.1b and J8.1.1b). The KS distance between the distributions illustrated in Figure~\ref{fig:B738_time_to_descend} is 0.158 and the Wasserstein distance is 31.8 seconds. This maps to Evidence E8.1.1c. Statistical distances that are computed using these metrics, while difficult to interpret, allow compliance against regulator-stipulated thresholds to be demonstrated in future. The reader can refer to \citet{hodgkin2025probabilisticsimulationaircraftdescent} for data on E8.1.1a and E8.1.1b (i.e.\ CAS and rate of descent).

\subsection*{Improved trajectory prediction}
A feature of the probabilistic trajectory predictor used in the Digital Twin is that it may be run in a deterministic mode which returns a most-likely TP, as calibrated against historical data. This is referred to as a `mean-mode' prediction as it is made using the mean of the data-driven correction to the thrust, drag, and speed terms which are added to BADA in the model. In mean-mode, the predictor can be used by agents and human controllers in place of BADA to perform TP. The mean-mode prediction gives the benefit of a TP model that is trained using data specific to a chosen airspace. It can therefore capture variations in aircraft performance that arise due to local airspace or operator procedures. This is in contrast to industry-standard models such as BADA, in which aircraft-specific parameters are calibrated against reference aircraft performance parameters \cite{nuic_paper}.

Property Claim 7 relates to the accuracy of these mean-mode predictions. Evidence to support sub-Property Claim 7.1.1 (not displayed in Figure~\ref{fig:assurance_real_world}) relates specifically to the accuracy of these predictions for descending B738s, and was generated through the following design of experiments:

\begin{enumerate}
    \item Initialise the probabilistic Digital Twin using aircraft positions, altitudes, clearances, and routes taken from the data 
    \item Generate simulated trajectories of each aircraft in the scenario using the mean-mode of the probabilistic trajectory predictor %This yields distributions of airspeed and rate of climb/descent (ROCD) for all aircraft in the scenario, which can be broken down by aircraft type
    \item Generate simulated trajectories of each aircraft in the scenario using BADA
    \item Compute the error between these sets of trajectories and the future states of the aircraft in the real-world traffic sample, broken down by aircraft type and phase of flight
\end{enumerate}

Evidence was generated using one week of data, using NATS' ARTAS and NERC systems to obtain radar and clearance information, respectively. The wind field was taken from the latest ECMWF ERA5 dataset for the same time period. Table~\ref{tab:real_world} displays the mean absolute error (MAE) between the trajectory samples generated from the two sets of simulations and the real-world data for descending aircraft. Green boxes indicate that the predictor in mean-mode achieved a lower MAE than BADA for both CAS and ROCD, by 27\% and 44\% respectively, a significant improvement. 

\begin{table}[h]
    \centering
    \caption{MAE for descending B738s on the week starting 2019-07-01.}
    \begin{tabular}{cccccc}
        \toprule
        Phase & Quantity & Digital Twin & BADA & Ratio \\
        \midrule
        %Cruise & CAS (kt) & \cellcolor[RGB]{179,255,179}15.38 & 15.94 & 0.96  \\
        %Climb & CAS (kt) & 17.43 &  \cellcolor[RGB]{255,179,179}15.20 & 1.15  \\
        Descent & CAS (kt) & \cellcolor[RGB]{179,255,179}16.77 & 22.95 & 0.73 \\
        %Climb & ROCD (ft/min) & \cellcolor[RGB]{179,255,179}313.69 & 319.97 & 0.98 \\
        Descent & ROCD (ft/min) &\cellcolor[RGB]{179,255,179}566.05 & 1009.22 & 0.56\\
        \bottomrule
    \end{tabular}
    \label{tab:real_world}
\end{table}

\subsection{Replication of NATS' ATCO assessment course}
Separate to the simulation of real-world operational airspace, the Digital Twin emulates the 31 formative  exercises of NATS' ATCO assessment course, as part of the competency-based framework described in \citet{Agent_validation_framework}. Therefore, the fidelity with which these exercises are replicated within the Digital Twin directly impacts its suitability as a platform for the development and assessment of AI agents for ATC. 

Replication of the formative exercises in the ATCO assessment course is the basis of Property Claim 6 for the argument over the accuracy and fidelity of the trajectory predictor (Figure \ref{fig:assurance_real_world}): `the trajectory predictor matches college data with sufficient accuracy and fidelity'. The sub-Property Claims in the assurance case relate to the fidelity with which individual exercises can be replicated in the Digital Twin. In this section, evidence to support these claims is presented. 

When running scenarios from NATS' ATCO training college, the Digital Twin generates aircraft dynamics using a specialised predictor subsystem tuned to replicate NATS' in-house simulator. Each scenario was run 32 times using the in-house simulator used in college assessments, with aircraft controlled by trainee ATCOs. This meant that multiple runs of each scenario were available, with differing sets of clearances issued during each run. The following design of experiments was used to generate evidence:
\begin{enumerate}
    \item For each run in the college data, initialise the Digital Twin using aircraft positions, altitudes, clearances, and routes taken from the original log file 
    \item Simulate the exercise using the Digital Twin, issuing clearances from the college log file at the same simulation time as in the college simulation
    \item After each radar refresh (every 6 seconds), compute the distance between aircraft in the Digital Twin simulation versus the same aircraft in the college data at that time
\end{enumerate}

\begin{figure}
    \centering
    \includegraphics[width=0.49\linewidth]{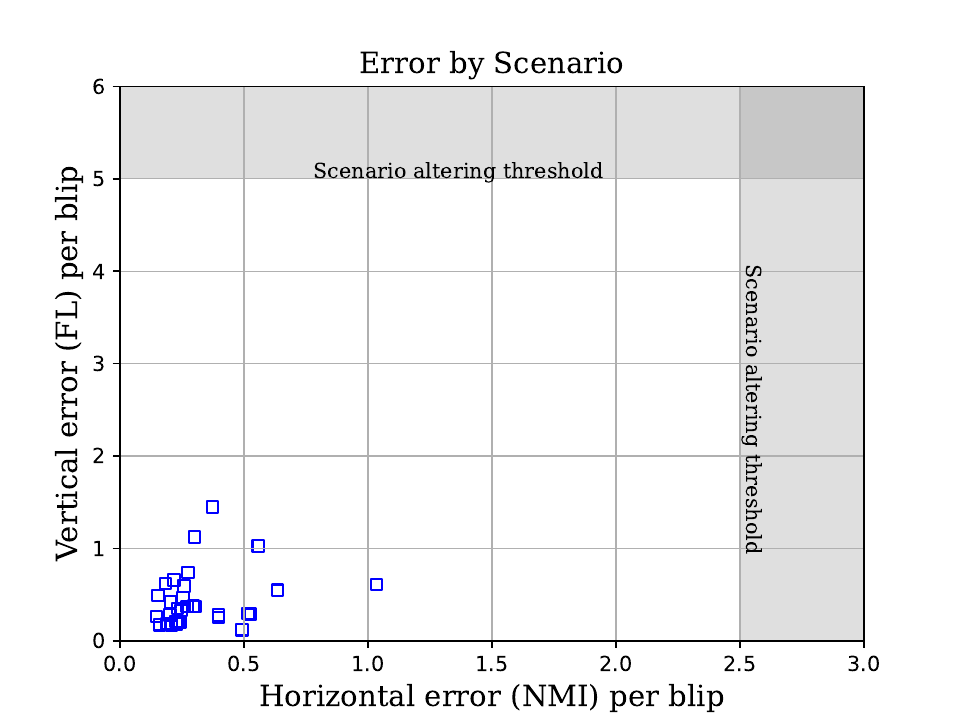}
    \caption{Average errors from replicating the 31 formative exercises in the ATCO assessment course in the Digital Twin, with regions above the instructor-informed thresholds shaded.}
    \label{fig:medway_scatter}
\end{figure}

Figure~\ref{fig:medway_scatter} is a scatter plot illustrating the mean error per formative exercise in the ATCO assessment course. Each point represents the error in a specific exercise averaged across all aircraft, blips in the simulation, and runs in the college data. Threshold values of 5~FL vertically and 2.5 nautical miles (NMI) laterally were suggested by instructors at the college as threshold values, below which mismatch between college data and Digital Twin simulations would have a negligible effect on the conflicts between aircraft in the scenarios, and therefore could be ignored. As can be seen in Figure~\ref{fig:medway_scatter} the average horizontal and vertical errors for all 31 formative exercises lie well below these thresholds, indicating that the Digital Twin faithfully reproduces the intended dynamics and difficulty profile of the training scenarios.

\begin{figure}
\centering
\begin{minipage}{0.45\linewidth}
   \centering
   \includegraphics[width=\linewidth]{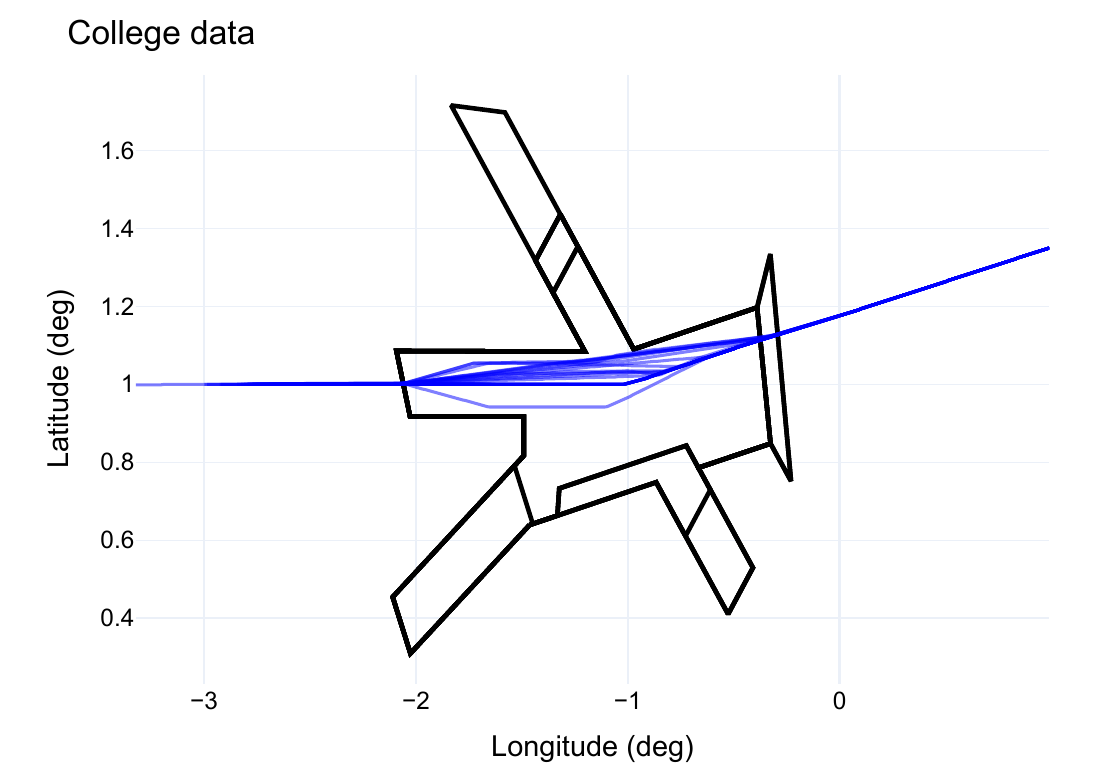}
\end{minipage}
\begin{minipage}{0.45\linewidth}
   \centering
   \includegraphics[width=\linewidth]{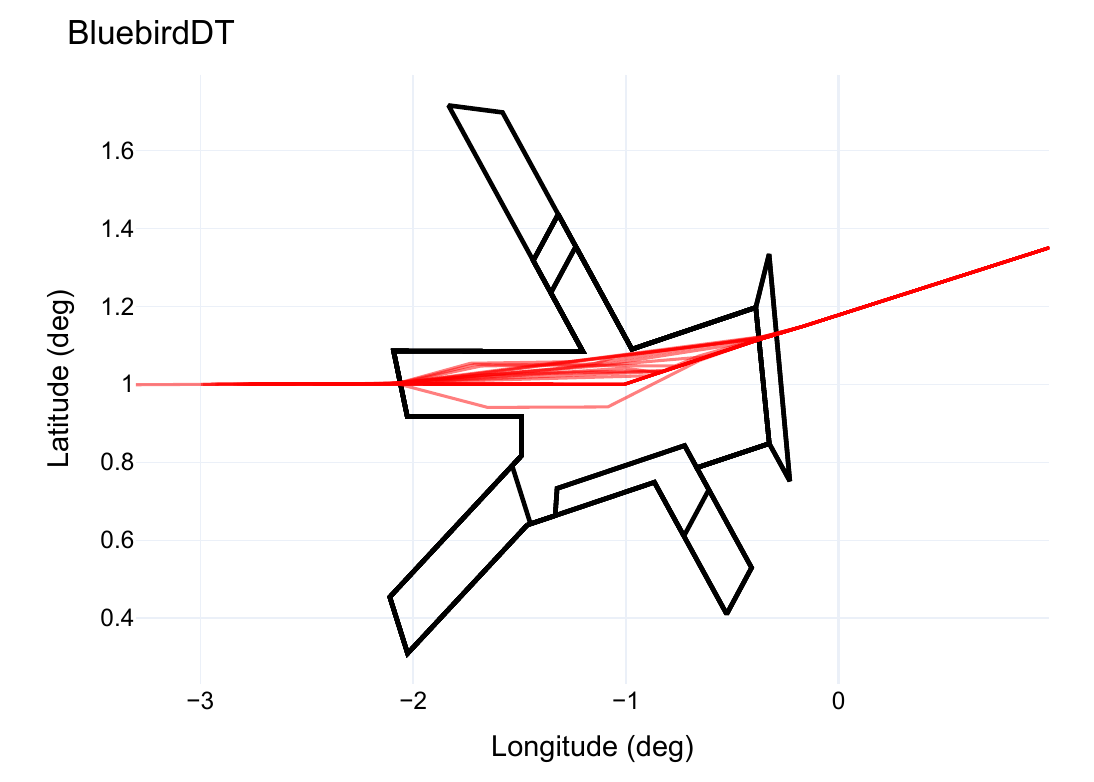}
\end{minipage}
\begin{minipage}{0.45\linewidth}
   \centering
   \includegraphics[width=\linewidth]{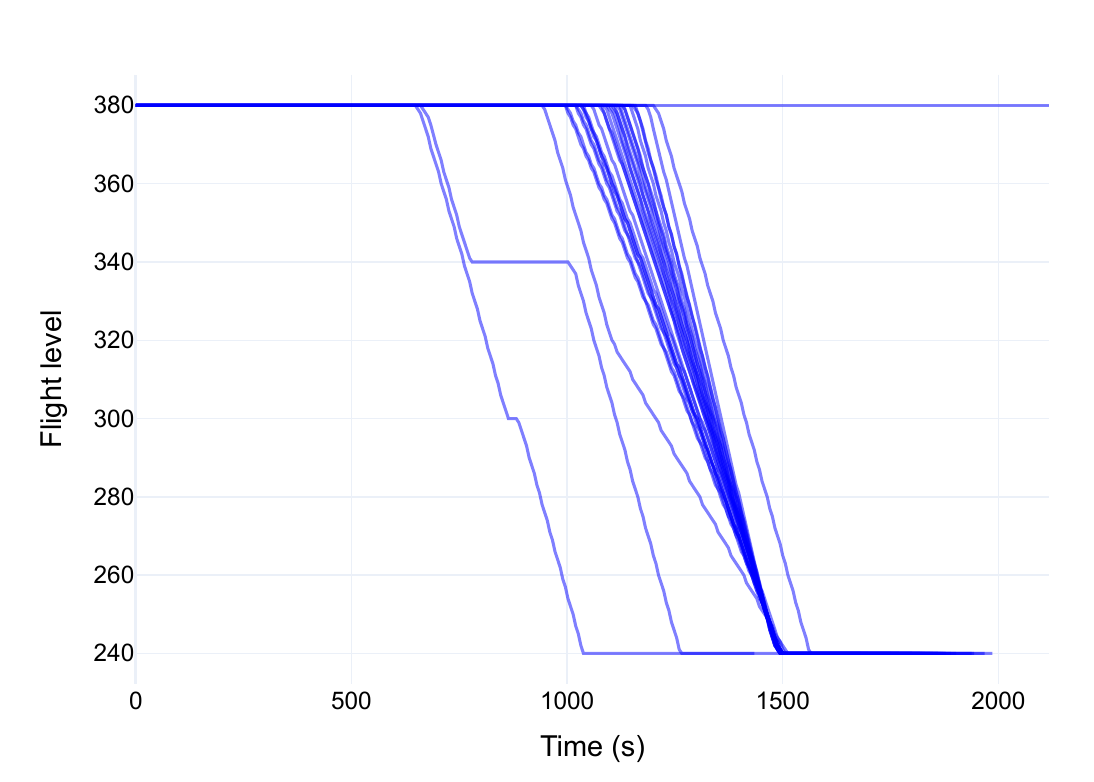}
\end{minipage}
\begin{minipage}{0.45\linewidth}
   \centering
   \includegraphics[width=\linewidth]{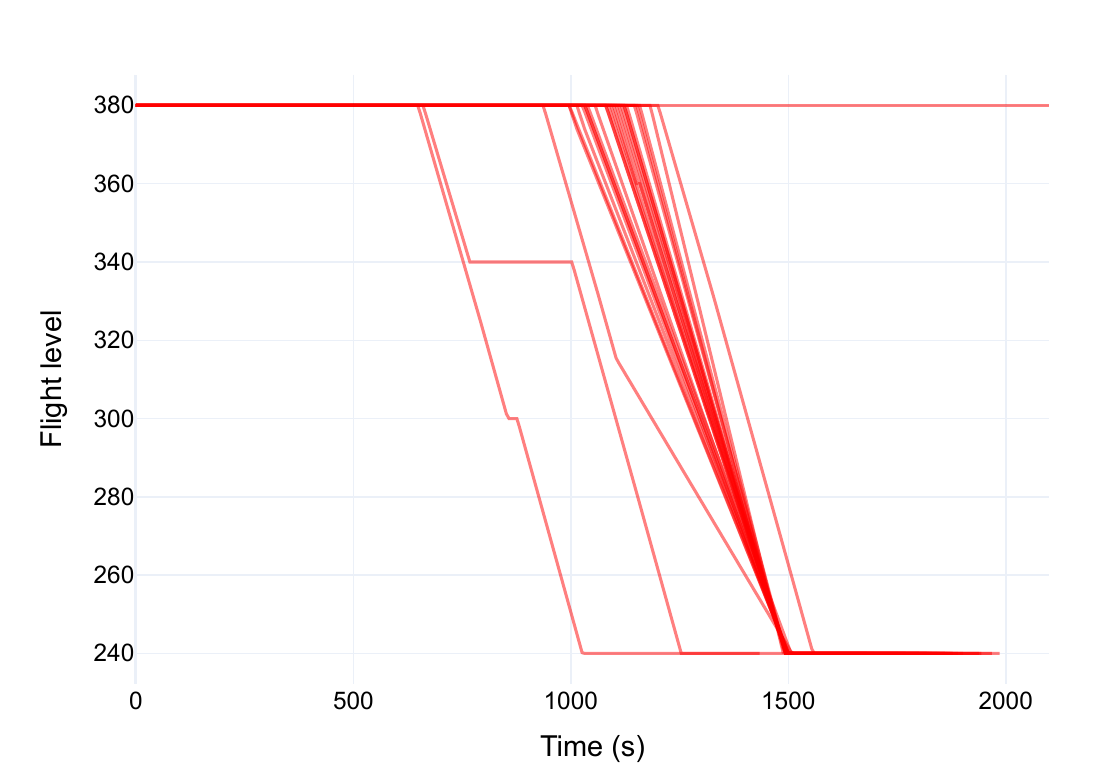}
\end{minipage}
\caption{Example of one aircraft in one exercise over the course of 28 runs. The left (blue) trajectories are those in the college data and the right (red) trajectories are those emulated by the trajectory predictor in the Digital Twin.}
\label{fig:medway_single_traj}
\end{figure}

\section{Future Work and Potential Applications} \label{sec:future}
This paper has focused on developing a probabilistic Digital Twin framework to train and evaluate AI agents for tactical control in en route airspace. Deploying such agents in live operations requires substantial technical validation and regulatory effort, which lies beyond the immediate scope of this work. Instead, our objective is to assess the maturity and limitations of these technologies to inform the roadmap toward higher levels of ATC automation. Nevertheless, the Digital Twin already replicates many characteristics of the real-world operational system and therefore enables several applications that bring benefits in the shorter term. This section outlines these near-term opportunities and future development paths.

The most significant barrier to operational use is the safety-critical nature of ATC, which appropriately demands exceptionally high levels of assurance. However, many impactful near-term applications exist outside direct operational control. 

\subsection*{AI agent development and formal evaluation}
The probabilistic Digital Twin has underpinned a series of trials executed using the Machine Basic Training (MBT) curriculum described by \citet{Agent_validation_framework}, evaluating a range of AI agents against a regulated standard of ATC performance. To date, five mature AI agents have been developed using the probabilistic Digital Twin, exploring a range of approaches from rules, search, and optimisation-based agents through to black box ML methods such as Reinforcement Learning \cite{hawk, mallard, RL_paper}. 

Two rounds of MBT trials were run in January and August of 2025, assessing the performance of the rules-based and optimisation-based agents, conducted in accordance with the MBT assessment framework to closely match the real-world process for assessing trainee ATCOs on their first simulator exams. Three 30-minute assessment runs were performed for each agent, with the results considered and moderated by a separate assessor to form a final judgement. Due in significant part to the high-fidelity environment provided by the Digital Twin, effective evidence could be gathered from these trials to guide agent development. This is evidenced by the progression of the rules-based agent from being rated \textit{``Unsatisfactory''} across all four assessed competencies, to being rated \textit{``Satisfactory''} in three out of four in the second round of trials, placing it close to a passing grade.

\subsection*{Prototype development and evaluation}
\begin{figure}
    \centering
    \includegraphics[width=0.8\linewidth]{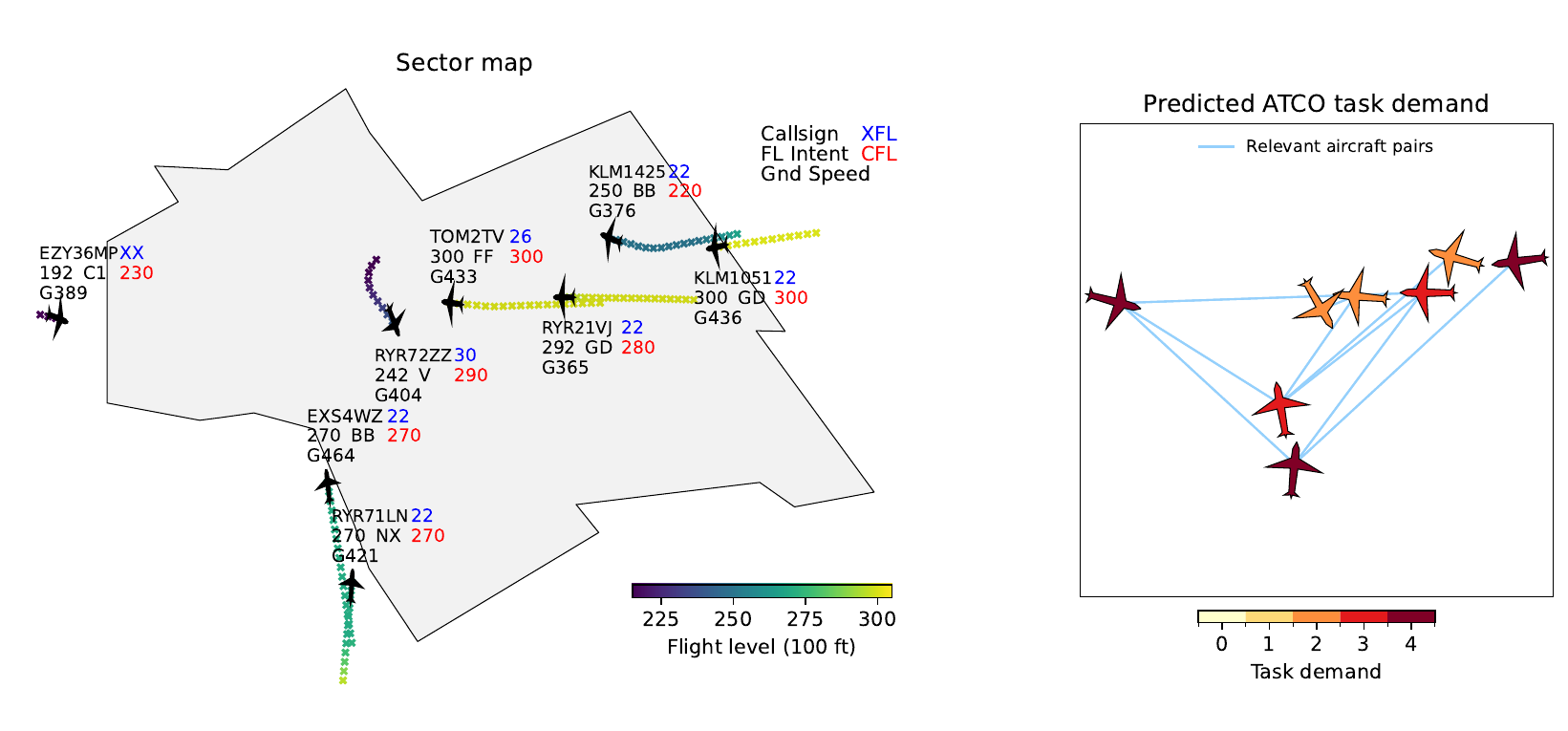}
    \caption{Output of a Graph Neural Network-based complexity analysis tool developed using the Digital Twin.}
    \label{fig:complexity_tool}
\end{figure}

\begin{figure}
    \centering
    \includegraphics[width=0.6\linewidth]{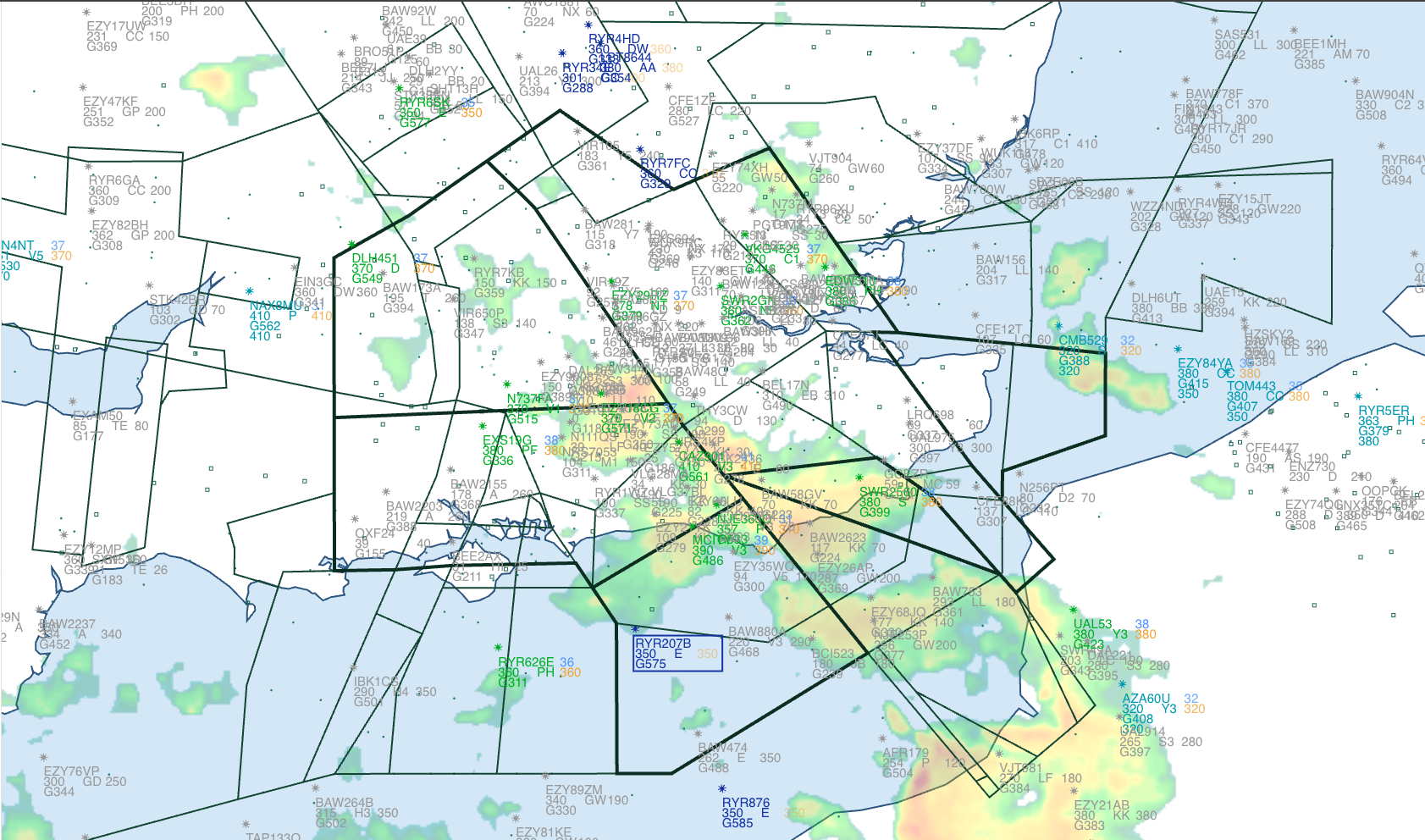}
    \caption{Precipitation overlay within the Digital Twin.}
    \label{fig:storm}
\end{figure}

The Digital Twin provides a high-fidelity replica of the operational environment, making it an ideal platform for prototyping and demonstrating new tools while retaining real-world fidelity. For example, \citet{complexity} developed a Graph Neural Network model to predict near-term ATCO task demand within a sector. Figure~\ref{fig:complexity_tool} shows an example of this tool's output, with the positions of aircraft in the sector on the left and the estimated cognitive load each aircraft entails for the ATCO on the right. Aircraft the tool flags as having direct relevance upon each other are joined with blue lines. Additionally, the platform has been used to explore a weather overlay system that fuses live traffic data with meteorological measurements such as precipitation rates, cloud coverage, water content, and convective available potential energy. These integrated data streams enable the development of data-driven models capable of estimating weather-induced delays or calculating the probability of aircraft diversions. Figure \ref{fig:storm} illustrates an example of this precipitation overlay.

\subsection*{Training and scenario generation}
The Digital Twin’s scenario generation, automated metrics, and replay functionalities offer valuable support for human ATCO training. To further enhance flexibility, recent work has utilised Large Language Models to convert textual prompts directly into executable scenarios, as detailed in \citet{gould2025airtrafficgenconfigurableairtraffic}.

\subsection*{Communication with the public and regulators}
\begin{figure}
    \centering
    \includegraphics[width=0.6\linewidth]{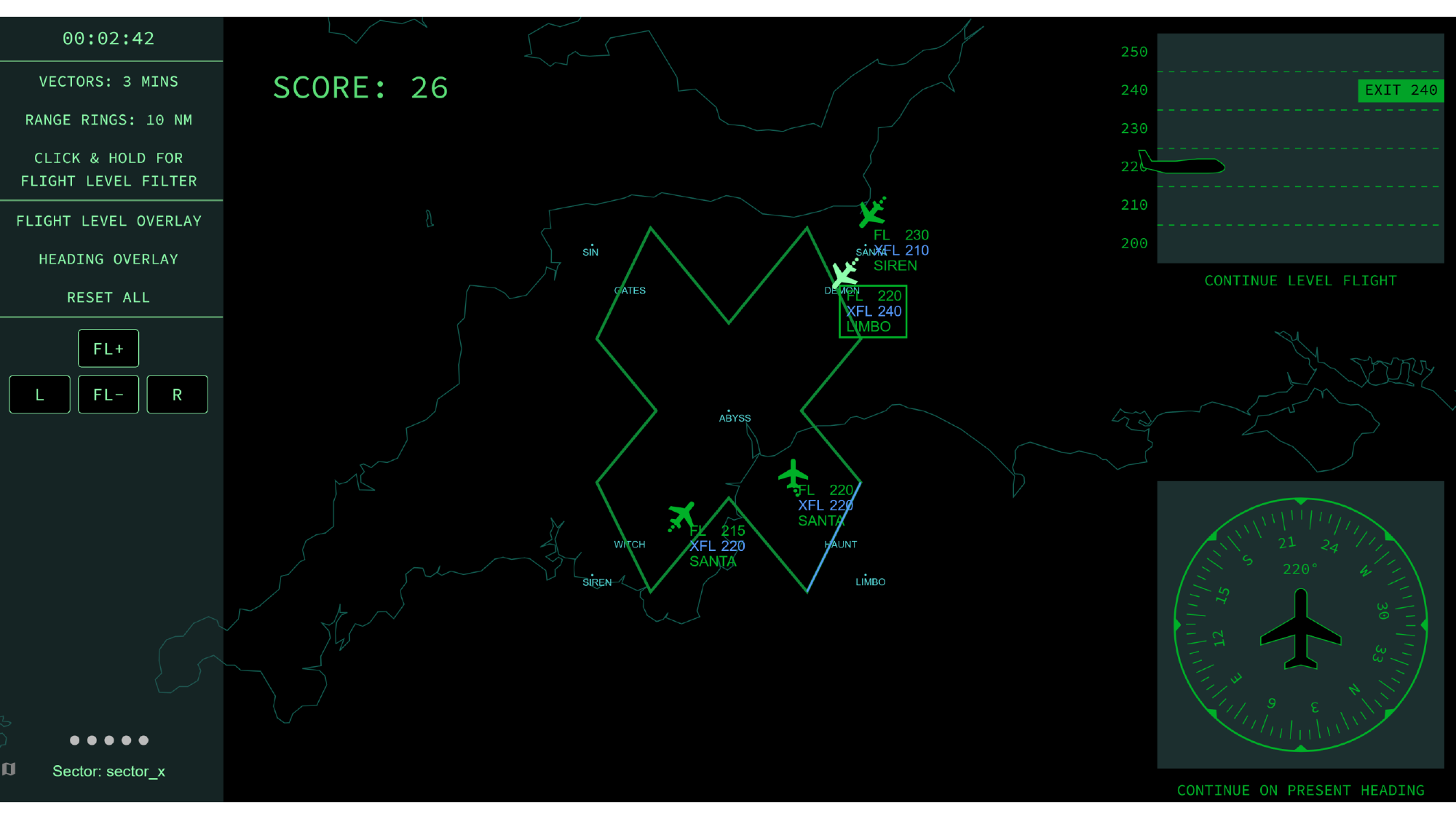}
    \caption{Screenshot of the public engagement game, hosted by the Digital Twin}
    \label{fig:game}
\end{figure}

A simplified, gamified iteration of the system, featuring keyboard-based aircraft control, has been deployed at outreach events to demonstrate ATC concepts to non-expert audiences (see Figure~\ref{fig:game}). Beyond public engagement, the Digital Twin provides a concrete implementation that facilitates meaningful discussions with regulators, thereby helping to inform and shape the regulatory framework for these emerging systems.

\subsection*{Digital Twin of Air Traffic Control}
While certified operational deployment of AI agents remains a significant long-term challenge, developing agents that approximate human controller performance in routine scenarios is substantially more achievable, with current agents already nearing this proficiency \cite{Agent_validation_framework}. When combined with the airspace Digital Twin, this composite system constitutes a ``Digital Twin of Air Traffic Control'' which enables a fast-time virtual representation of the complete ATC task, rather than solely the physical environment. Such a platform facilitates the rapid, parallel exploration of new procedures, airspace designs, and routing strategies, filtering promising concepts before committing to costly human-in-the-loop trials. Furthermore, it enhances flow and capacity management by jointly modelling the interaction between aircraft dynamics and controller behaviour.

\subsection*{Future Development}
While the Digital Twin reproduces the operational environment with high-fidelity, specific limitations constrain its current scope. At present, detailed modelling is restricted to the en route LACC airspace, while the more dynamic lower airspace of London Terminal Control Centre (LTCC) is represented in a simplified form. Furthermore, the probabilistic trajectory prediction model relies on historical data and does not yet capture aircraft-specific performance variations inferred from live radar tracks, which limits short-term predictive accuracy for individual flights. Finally, the existing assurance case addresses the Digital Twin's fidelity rather than a full operational safety case or the joint assurance challenges introduced by AI agents interacting with the system. These constraints motivate the development directions outlined below.

Future work will expand both the fidelity and geographic coverage of the Digital Twin. The primary enhancement will incorporate LTCC and simplified airport models, enabling an end-to-end representation of one of the world's most complex and capacity-constrained airspaces. This will also entail expanding the set of clearances supported in the Digital Twin, in order to accurately model the larger instruction set utilised in this area of the operation. Subsequent phases will cover the remainder of UK airspace and the UK-controlled eastern North Atlantic. Additionally, scenario generation capabilities will be extended to automate the creation of unusual or emergency events, such as incursions, radio failures, and cabin decompressions, which currently require manual configuration.

Advances in probabilistic TP constitute a parallel development stream. Although the existing type-specific model exceeds BADA fidelity, it does not account for individual aircraft performance differences. Ongoing research aims to assimilate per-aircraft data to improve short-term predictive accuracy, utilising methods such as those described in our recent work~\citep{Assimilation_paper}.%\ben{Perhaps mention expanding the action space? We currently have quite a restricted set compared to operations, and the complexity of implementing level by is a sign of things to come. Maybe mention the trade off between fidelity and usability for research.}

Finally, an open-source release of the Digital Twin is planned for April 2026. This public version will provide a subset of the full system's capabilities, excluding proprietary or licensed components, while still offering a high-value platform for research, experimentation, and innovation to the wider community.

\FloatBarrier

\section{Conclusion}\label{sec:conclusion}

This paper presents what is, to the authors' knowledge, the first probabilistic Digital Twin of operational en route airspace, covering the London Area Control environment, and designed to support the development and evaluation of AI agents for Air Traffic Control. The system integrates high-fidelity, AIRAC-aligned models of airspace structure and sectorisation with both historical and live operational data sources, including surveillance, flight plans, clearances, sector configurations, and meteorological information. Aircraft motion is modelled either through direct trajectory replay or via a probabilistic, physics-informed machine learning generator, which effectively captures the uncertainties intrinsic to real-world operations. This realises this the first contribution of this work, a probabilistic Digital Twin of operational en route airspace, rooted in real-world data and physics-based constraints.

Beyond physical reproduction, the system offers a comprehensive framework for AI agent development, training and assessment. Key capabilities include fast-time execution for rapid experimentation, a set of quantitative performance metrics, and a standardised ``gym'' interface for benchmarking. A human-in-the-loop capability allows qualified controllers to observe, query, and intervene in agent-led scenarios, while the integration of formative exercises from NATS' trainee assessment course enables structured, competency-based evaluation of AI agents under controlled conditions. Together, these elements deliver the second contribution, a unified framework that supports the full AI agent lifecycle from prototyping through to systematic evaluation.

Given the safety-critical nature of ATC and the novelty of ML-enabled probabilistic Digital Twins, assurance is a central focus of this work. Using the Trustworthy and Ethical Assurance platform, we developed a Goal Structured Notation assurance case addressing the accuracy and fidelity of the twin. Within this framework, we presented quantitative evidence demonstrating that the probabilistic trajectory predictor reproduces real-world aircraft behaviour with greater accuracy than a baseline BADA model, and that the replicated exercises preserve operational intent to a degree sufficient for rigorous training and evaluation. This constitutes this work's third contribution, a quantitative validation of the Digital Twin, embedded within a structured assurance argument.

The Digital Twin already enables applications in AI agent development, prototype evaluation, human-machine teaming studies, and ATCO training. It provides an example of Digital Twinning technology in ATM that can be used to inform engagement with regulators as the guidance for operationalising this technology evolves. Ongoing and future work will expand the system's scope to include terminal control and the wider UK airspace, enhance the trajectory model via live data assimilation, and broaden community access through an open-source release.

In summary, the probabilistic Digital Twin presented here delivers a realistic, data-driven, and quantitatively assured environment for exploring higher levels of automation in ATC. By enabling AI agents to be trained and evaluated under realistic uncertainty, with operationally credible traffic and interactive human oversight, it provides a practical foundation for understanding when, where, and how such agents could safely and effectively contribute to future Air Traffic Management systems.
%and it serves as a tangible technological demonstrator to facilitate evidence-based engagement with regulators

%\section*{Appendix}

\section*{Funding Sources}
The work described in this paper is funded by the grant “EP/V056522/1: Advancing Probabilistic Machine Learning to Deliver Safer, More Efficient and Predictable Air Traffic Control” (aka Project Bluebird), an EPSRC Prosperity Partnership between NATS, The Alan Turing Institute, and the University of Exeter. 

\section*{Acknowledgments}
We would like to thank the Research Engineering Group at The Alan Turing Institute, the Research Software Engineers Group at the University of Exeter, and the software engineers at NATS who have been an important part of the Project Bluebird team, delivering the Digital Twin codebase.

\bibliography{starling}

@manual{argos,
  title  = {{ARGOS Factsheet}},
  author = {{EUROCONTROL}},
  year   = {2023},
  url    = {https://www.eurocontrol.int/publication/argos-factsheet}
}

@manual{easr_2022,
  title  = {{2022 European Aviation Environmental Report}},
  author = {{European Union Aviation Safety Agency (EASA)}},
  year   = {2022},
  url    = {https://www.easa.europa.eu/eco/eaer}
}

@manual{eurocontrol_artas_web,
  author       = {{EUROCONTROL}},
  title        = {{ARTAS}: Air traffic management surveillance tracker and server},
  year         = {n.d.},
  url = {https://www.eurocontrol.int/product/artas},
}

@manual{faa,
  author  = {{Federal Aviation Administration}},
  title   = {Next Generation Air Transportation System (NextGen)},
  year    = {2023},
  url     = {https://www.faa.gov/nextgen},
  urldate = {24-03-2023}
}

@article{hodgkin2025probabilisticsimulationaircraftdescent,
  title   = {Probabilistic Simulation of Aircraft Descent via a Physics-Informed Machine Learning Approach},
  author  = {Amy Hodgkin and Nick Pepper and Marc Thomas},
  year    = {2025},
  journal = {arXiv},
  url     = {https://arxiv.org/abs/2504.02529}
}

@manual{jetzero,
  author  = {{Department for Transport (UK)}},
  title   = {Jet Zero Strategy: Delivering net zero aviation by 2050},
  year    = {2022},
  url     = {https://www.gov.uk/government/publications/jet-zero-strategy-delivering-net-zero-aviation-by-2050},
  urldate = {16-05-2025}
}

@article{kapteyn2021probabilistic,
  title     = {A probabilistic graphical model foundation for enabling predictive digital twins at scale},
  author    = {Kapteyn, Michael G and Pretorius, Jacob VR and Willcox, Karen E},
  journal   = {Nature Computational Science},
  volume    = {1},
  number    = {5},
  pages     = {337--347},
  year      = {2021},
  publisher = {Nature Publishing Group US New York},
  url = {https://doi.org/10.1038/s43588-021-00069-0}
}

@inproceedings{kim_aerodts,
  author    = {Kim, Myeung Un},
  booktitle = {2022 13th International Conference on Information and Communication Technology Convergence (ICTC)},
  title     = {A Survey on Digital Twin in Aerospace in the New Space Era},
  year      = {2022},
  volume    = {},
  number    = {},
  page      = {1735--1737},
  url = {https://doi.org/10.1109/ICTC55196.2022.9952929}
}

@article{li2017dynamic,
  title     = {Dynamic Bayesian network for aircraft wing health monitoring digital twin},
  author    = {Li, Chenzhao and Mahadevan, Sankaran and Ling, You and Choze, Sergio and Wang, Liping},
  journal   = {Aiaa Journal},
  volume    = {55},
  number    = {3},
  pages     = {930--941},
  year      = {2017},
  publisher = {American Institute of Aeronautics and Astronautics},
  url = {https://doi.org/10.2514/1.J055201}
}

@techreport{nats2012fuel,
  title       = {NATS Fuel Efficiency Metric},
  author      = {Chris Nutt},
  institution = {NATS},
  year        = {2012},
  url         = {https://www.nats.aero/wp-content/uploads/2012/03/fuelEfficiencyMetric.pdf}
}

@article{nuic2010user,
  title   = {User manual for the Base of Aircraft Data (BADA) revision 3.10},
  author  = {Nuic, Angela},
  journal = {Atmosphere},
  volume  = {2010},
  pages   = {001},
  year    = {2010},
  url     =  {http://maartenuijtdehaag.com/bada310-user-manual.pdf}
}

@article{nuic_paper,
author = {Nuic, Angela and Poles, Damir and Mouillet, Vincent},
title = {BADA: An advanced aircraft performance model for present and future ATM systems},
journal = {International Journal of Adaptive Control and Signal Processing},
volume = {24},
number = {10},
pages = {850-866},
url = {https://onlinelibrary.wiley.com/doi/abs/10.1002/acs.1176},
year = {2010}
}

@article{Pepper_DataBADA,
  author  = {Pepper, Nick and Thomas, Marc},
  title   = {Learning Generative Models for Climbing Aircraft from Radar Data},
  journal = {Journal of Aerospace Information Systems},
  volume  = {21},
  number  = {6},
  pages   = {474--481},
  year    = {2024},
  url = {https://doi.org/10.2514/1.I011359}
}

@article{pepper4984556probabilistic,
  title   = {A sector-specific probabilistic approach for 4D aircraft trajectory generation},
  journal = {Transportation Research Part C: Emerging Technologies},
  volume  = {179},
  pages   = {105291},
  year    = {2025},
  issn    = {0968-090X},
  author  = {Nick Pepper and George {De Ath} and Ben Carvell and Amy Hodgkin and Tim Dodwell and Marc Thomas and Richard Everson},
  url = {https://doi.org/10.1016/j.trc.2025.105291}
}

@inproceedings{prob_rotorcraft_DT,
  author    = {Sankaran Mahadevan and William Sisson and Pranav Karve},
  title     = {Probabilistic Digital Twin for Damage-Aware Rotorcraft Maneuvering},
  booktitle = {AIAA SCITECH 2025 Forum},
  chapter   = {},
  pages     = {},
  year      = 2025,
  url = {https://doi.org/10.2514/6.2025-2817}
}

@manual{sesar,
  author = {{SESAR Joint Undertaking}},
  title  = {Delivering the Digital European Sky},
  year   = {2021},
  url    = {https://www.sesarju.eu/}
}

@manual{Sesar_sol,
  author = {{SESAR Joint Undertaking}},
  title  = {{SESAR} solutions catalogue (Fourth edition)},
  year   = {2021},
  url    = {https://www.sesarju.eu/activities-solutions}
}

@article{gould2025airtrafficgenconfigurableairtraffic,
  title         = {AirTrafficGen: Configurable Air Traffic Scenario Generation with Large Language Models},
  author        = {Dewi Sid William Gould and George {De Ath} and Ben Carvell and Nick Pepper},
  year          = {2025},
  eprint        = {2508.02269},
  journal = {arXiv},
  url = {https://arxiv.org/abs/2508.02269}
}

@inproceedings{complexity,
  title     = {Air traffic controller task demand via graph neural networks: An interpretable approach to airspace complexity},
  author    = {Henderson, Edward and Gould, Dewi and {De Ath}, George and Everson, Richard and Pepper, Nick},
  booktitle = {AIAA AVIATION FORUM AND ASCEND 2025},
  pages     = {3590},
  year      = {2025},
  url = {https://doi.org/10.2514/6.2025-3590}
}

@inproceedings{hoekstra2016bluesky,
  title        = {Bluesky ATC simulator project: an open data and open source approach},
  author       = {Hoekstra, Jacco M and Ellerbroek, Joost},
  booktitle    = {Proceedings of the 7th international conference on research in air transportation},
  volume       = {131},
  pages        = {132},
  year         = {2016},
  organization = {FAA/Eurocontrol Washington, DC, USA},
  url = {https://pure.tudelft.nl/ws/files/10083831/Hoekstra_BlueSky_project.pdf}
}

@article{loft2004atc,
  title     = {ATC-lab: An air traffic control simulator for the laboratory},
  author    = {Loft, Shayne and Hill, Andrew and Neal, Andrew and Humphreys, Michael and Yeo, Gillian},
  journal   = {Behavior Research Methods, Instruments, \& Computers},
  volume    = {36},
  number    = {2},
  pages     = {331--338},
  year      = {2004},
  publisher = {Springer},
  url = {https://doi.org/10.3758/BF03195579}
}

@article{bilimoria2001facet,
  title     = {FACET: Future ATM concepts evaluation tool},
  author    = {Bilimoria, Karl D and Sridhar, Banavar and Grabbe, Shon R and Chatterji, Gano B and Sheth, Kapil S},
  journal   = {Air Traffic Control Quarterly},
  volume    = {9},
  number    = {1},
  pages     = {1--20},
  year      = {2001},
  publisher = {American Institute of Aeronautics and Astronautics, Inc.},
  url = {https://doi.org/10.2514/atcq.9.1.1}
}

@manual{vatsim,
  author  = {{VATSIM}},
  title   = {{The International Online Aviation Network}},
  year    = {n.d.},
  url     = {https://vatsim.net/},
  urldate = {2025-10-20}
}

@manual{eurocontrol_escape,
  author  = {{EUROCONTROL}},
  title   = {{EUROCONTROL simulation capabilities and platform for experimentation}},
  year    = {n.d.},
  url     = {https://www.eurocontrol.int/simulator/escape},
  urldate = {2025-10-20}
}

@manual{eurocontrol_edep,
  author  = {{EUROCONTROL}},
  title   = {{Early demonstration and evaluation platform}},
  year    = {n.d.},
  url     = {https://www.eurocontrol.int/simulator/edep},
  urldate = {2025-10-20}
}

@inproceedings{molina2014agent,
  title        = {Agent-based modeling and simulation for the design of the future european air traffic management system: The experience of cassiopeia},
  author       = {Molina, Martin and Carrasco, Sergio and Martin, Jorge},
  booktitle    = {International Conference on Practical Applications of Agents and Multi-Agent Systems},
  pages        = {22--33},
  year         = {2014},
  organization = {Springer},
  url = {https://doi.org/10.1007/978-3-319-07767-3_3}
}

@manual{openScope,
  author  = {{openScope}},
  title   = {{Air Traffic Control Simulator}},
  year    = {2022},
  url     = {www.openscope.io },
  urldate = {2025-10-20}
}

@manual{atc_sim,
  author  = {{ATC-SIM}},
  title   = {{A browser-based air traffic control simulator game}},
  year    = {n.d.},
  url     = {https://atc-sim.com/},
  urldate = {2025-10-20}
}

@inproceedings{munoz2015formal,
  title        = {Formal methods in air traffic management: The case of unmanned aircraft systems (invited lecture)},
  author       = {Munoz, C{\'e}sar A},
  booktitle    = {International Colloquium on Theoretical Aspects of Computing},
  pages        = {58--62},
  year         = {2015},
  organization = {Springer}
}

@article{AV_scengen_review,
  author         = {Cai, Jinkang and Deng, Weiwen and Guang, Haoran and Wang, Ying and Li, Jiangkun and Ding, Juan},
  title          = {A Survey on Data-Driven Scenario Generation for Automated Vehicle Testing},
  journal        = {Machines},
  volume         = {10},
  year           = {2022},
  number         = {11},
  article-number = {1101},
  issn           = {2075-1702},
  url = {https://doi.org/10.3390/machines10111101}
}

@article{self_driving_cars,
  author  = {Ding, Wenhao and Xu, Chejian and Arief, Mansur and Lin, Haohong and Li, Bo and Zhao, Ding},
  journal = {IEEE Transactions on Intelligent Transportation Systems},
  title   = {A Survey on Safety-Critical Driving Scenario Generation—A Methodological Perspective},
  year    = {2023},
  volume  = {24},
  number  = {7},
  pages   = {6971--6988},
  url = {https://ieeexplore.ieee.org/document/10089194}
}

@article{katsoulakis2024digital,
  title     = {Digital twins for health: a scoping review},
  author    = {Katsoulakis, Evangelia and Wang, Qi and Wu, Huanmei and Shahriyari, Leili and Fletcher, Richard and Liu, Jinwei and Achenie, Luke and Liu, Hongfang and Jackson, Pamela and Xiao, Ying and others},
  journal   = {NPJ digital medicine},
  volume    = {7},
  number    = {1},
  pages     = {77},
  year      = {2024},
  publisher = {Nature Publishing Group UK London},
  url = {https://doi.org/10.1038/s41746-024-01073-0}
}

@article{yu2022energy,
  title     = {Energy digital twin technology for industrial energy management: Classification, challenges and future},
  author    = {Yu, Wei and Patros, Panos and Young, Brent and Klinac, Elsa and Walmsley, Timothy Gordon},
  journal   = {Renewable and Sustainable Energy Reviews},
  volume    = {161},
  pages     = {112407},
  year      = {2022},
  publisher = {Elsevier},
  url = {https://doi.org/10.1016/j.rser.2022.112407}
}

@article{kuvsic2023digital,
  title     = {A digital twin in transportation: Real-time synergy of traffic data streams and simulation for virtualizing motorway dynamics},
  author    = {Ku{\v{s}}i{\'c}, Kre{\v{s}}imir and Schumann, Ren{\'e} and Ivanjko, Edouard},
  journal   = {Advanced Engineering Informatics},
  volume    = {55},
  pages     = {101858},
  year      = {2023},
  publisher = {Elsevier},
  url = {https://doi.org/10.1016/j.aei.2022.101858}
}

@article{wu2022digital,
  title     = {Digital twins and artificial intelligence in transportation infrastructure: Classification, application, and future research directions},
  author    = {Wu, Jingyi and Wang, Xiao and Dang, Yukun and Lv, Zhihan},
  journal   = {Computers and Electrical Engineering},
  volume    = {101},
  pages     = {107983},
  year      = {2022},
  publisher = {Elsevier}
}

@article{michalek2009identification,
  issn      = {00411655, 15265447},
  author    = {Diana Michalek Pfeil and Hamsa Balakrishnan},
  journal   = {Transportation Science},
  number    = {1},
  pages     = {56--73},
  publisher = {INFORMS},
  title     = {Identification of Robust Terminal-Area Routes in Convective Weather},
  volume    = {46},
  year      = {2012},
  url       = {http://www.jstor.org/stable/41432825},
}

@article{pyrgiotis2013modelling,
  title     = {Modelling delay propagation within an airport network},
  author    = {Pyrgiotis, Nikolas and Malone, Kerry M and Odoni, Amedeo},
  journal   = {Transportation Research Part C: Emerging Technologies},
  volume    = {27},
  pages     = {60--75},
  year      = {2013},
  publisher = {Elsevier},
  url = {https://doi.org/10.1016/j.trc.2011.05.017}
}

@inproceedings{sridhar2009modeling,
  title     = {Modeling flight delays and cancellations at the national, regional and airport levels in the United States},
  author    = {Sridhar, Banavar and Wang, Yao and Klein, Alexander and Jehlen, Richard},
  booktitle = {8th USA/Europe ATM R\&D Seminar, Napa, California (USA)},
  year      = {2009},
  url={https://api.semanticscholar.org/CorpusID:33815016}

}

@inbook{route_uncertainty2,
  author    = {Craig Wanke and Michael Callaham and Daniel Greenbaum and Anthony Masalonis},
  title     = {Measuring Uncertainty in Airspace Demand Predictions for Traffic Flow Management Applications},
  booktitle = {AIAA Guidance, Navigation, and Control Conference and Exhibit},
  chapter   = {},
  pages     = {},
  url = {https://doi.org/10.2514/6.2003-5708}, 
  year      = 2012
}

@inproceedings{route_uncertainty1,
  author    = {Sandamali, Gammana Guruge Nadeesha and Su, Rong and Zhang, Yicheng and Li, Qing},
  booktitle = {2017 13th IEEE International Conference on Control \& Automation (ICCA)},
  title     = {Flight routing and scheduling with departure uncertainties in air traffic flow management},
  year      = {2017},
  volume    = {},
  number    = {},
  pages     = {301-306},
  url = {https://ieeexplore.ieee.org/document/8003077}
}

@inproceedings{matthews2010assessment,
  title     = {Assessment and interpretation of en route weather avoidance fields from the convective weather avoidance model},
  author    = {Matthews, Michael and DeLaura, Rich},
  booktitle = {10th AIAA Aviation Technology, Integration, and Operations (ATIO) Conference},
  pages     = {9160},
  year      = {2010}
}

@manual{atc_emergencies,
  author  = {{EUROPEAN ORGANISATION FOR THE SAFETY OF AIR NAVIGATION}},
  title   = {{Guidelines for Controller Training in the Handling of Unusual/Emergency Situations}},
  year    = {2003},
  url     = {https://skybrary.aero/sites/default/files/bookshelf/15.pdf},
  urldate = {2025-10-20}
}

@article{sun2018aircraft,
  title     = {Aircraft initial mass estimation using Bayesian inference method},
  author    = {Sun, Junzi and Ellerbroek, Joost and Hoekstra, Jacco M},
  journal   = {Transportation Research Part C: Emerging Technologies},
  volume    = {90},
  pages     = {59--73},
  year      = {2018},
  publisher = {Elsevier},
  url = {https://doi.org/10.1016/j.trc.2018.02.022}
}

@article{aerospace9020091,
  author  = {Zeng, Weili and Chu, Xiao and Xu, Zhengfeng and Liu, Yan and Quan, Zhibin},
  title   = {Aircraft 4D Trajectory Prediction in Civil Aviation: A Review},
  journal = {Aerospace},
  volume  = {9},
  year    = {2022},
  number  = {2},
  issn    = {2226-4310},
  url     = {https://doi.org/10.3390/aerospace9020091}
}

@article{met_sensitivity_tp,
  author  = {Jackson, Michael R. and Zhao, Yiyuan J. and Slattery, Rhonda A.},
  title   = {Sensitivity of Trajectory Prediction in Air Traffic Management},
  journal = {Journal of Guidance, Control, and Dynamics},
  volume  = {22},
  number  = {2},
  pages   = {219-228},
  year    = {1999},
  url     = {https://doi.org/10.2514/2.4398}
}

@inproceedings{dt_validation,
  author    = {Hua, Edward Y. and Lazarova-Molnar, Sanja and Francis, Deena P.},
  booktitle = {2022 Winter Simulation Conference (WSC)},
  title     = {Validation of Digital Twins: Challenges and Opportunities},
  year      = {2022},
  volume    = {},
  number    = {},
  pages     = {2900-2911},
  url = {https://doi.org/10.1109/WSC57314.2022.10015420}
}

@techreport{burr2024,
  title       = {Trustworthy and {{Ethical Assurance}} of {{Digital Twins}}: {{Putting}} the {{Gemini Principles}} into {{Practice}}},
  shorttitle  = {Trustworthy and {{Ethical Assurance}} of {{Digital Twins}}},
  author      = {Burr, Christopher and Anderson, Justin and Arana, Sophie and Block, Daniel and Byrne, James and Goodman, Ryan and {Gavidia-Calderon}, Carlos and Habli, Ibrahim and Moreira, Nury and Niederer, Steven and Polo, Nuala and Wagg, David},
  year        = {2024},
  month       = dec,
  institution = {Alan Turing Institute},
  url = {https://zenodo.org/records/14259252},
  urldate     = {2025-05-21},
  langid      = {english},
url = {}
}

@manual{easa_guidelines_2024,
  author = {{European Union Aviation Safety Agency}},
  title  = {EASA Artificial Intelligence (AI) Concept Paper: Guidance for Level 1 \& 2 machine learning application},
  year   = {2024},
  url    = {https://www.easa.europa.eu/en/document-library/general-publications/easa-artificial-intelligence-concept-paper-issue-2}
}

@manual{easa_atco_training,
  author = {{European Union Aviation Safety Agency}},
  title  = {Acceptable Means of Compliance (AMC) and Guidance Material (GM) to Part ATCO.OR: Requirements for air traffic controller training organisations and aero-medical centres},
  year   = {2015},
  url    = {https://www.easa.europa.eu/sites/default/files/dfu/annex_iii_to_ed_decision_2015-010-r.pdf}
}

@manual{caa_cap2970,
  author  = {CAA},
  title   = {CAP2970: Building Trust in AI},
  year    = {2024},
  url     = {https://www.caa.co.uk/publication/download/21120},
  urldate = {21-05-2025}
}

@article{faa_roadmap,
  author = {{Federal Aviation Administration}},
  title  = {Roadmap for Artificial Intelligence Safety Assurance},
  year   = {2024},
  url    = {https://www.faa.gov/aircraft/air_cert/step/roadmap_for_AI_safety_assurance}
}

@book{nasem2024,
  title     = {Foundational Research Gaps and Future Directions for Digital Twins},
  author    = {{National Academies of Sciences, Engineering, {and} Medicine.}},
  year      = {2024},
  address   = {Washington, DC},
  publisher = {The National Academies Press},
}

@article{wagg2020digital,
  title     = {Digital twins: state-of-the-art and future directions for modeling and simulation in engineering dynamics applications},
  author    = {Wagg, DJ and Worden, Keith and Barthorpe, RJ and Gardner, Paul},
  journal   = {ASCE-ASME Journal of Risk and Uncertainty in Engineering Systems, Part B: Mechanical Engineering},
  volume    = {6},
  number    = {3},
  pages     = {030901},
  year      = {2020},
  publisher = {American Society of Mechanical Engineers},
  url = {https://doi.org/10.1115/1.4046739}
}

@article{ERA5,
  author   = {Hersbach, Hans and Bell, Bill and Berrisford, Paul and Hirahara, Shoji and Horányi, András and Muñoz-Sabater, Joaquín and Nicolas, Julien and Peubey, Carole and Radu, Raluca and Schepers, Dinand and Simmons, Adrian and Soci, Cornel and Abdalla, Saleh and Abellan, Xavier and Balsamo, Gianpaolo and Bechtold, Peter and Biavati, Gionata and Bidlot, Jean and Bonavita, Massimo and De Chiara, Giovanna and Dahlgren, Per and Dee, Dick and Diamantakis, Michail and Dragani, Rossana and Flemming, Johannes and Forbes, Richard and Fuentes, Manuel and Geer, Alan and Haimberger, Leo and Healy, Sean and Hogan, Robin J. and Hólm, Elías and Janisková, Marta and Keeley, Sarah and Laloyaux, Patrick and Lopez, Philippe and Lupu, Cristina and Radnoti, Gabor and de Rosnay, Patricia and Rozum, Iryna and Vamborg, Freja and Villaume, Sebastien and Thépaut, Jean-Noël},
  title    = {The ERA5 global reanalysis},
  journal  = {Quarterly Journal of the Royal Meteorological Society},
  volume   = {146},
  number   = {730},
  pages    = {1999-2049},
  year     = {2020},
  url = { https://doi.org/10.1002/qj.3803}
}

@article{drones7070484,
  author  = {Souanef, Toufik and Al-Rubaye, Saba and Tsourdos, Antonios and Ayo, Samuel and Panagiotakopoulos, Dimitrios},
  title   = {Digital Twin Development for the Airspace of the Future},
  journal = {Drones},
  volume  = {7},
  year    = {2023},
  number  = {7},
  issn    = {2504-446X},
  url = {https://doi.org/10.3390/drones7070484}
}

@misc{openai_gym,
  author = {Greg Brockman and Vicki Cheung and Ludwig Pettersson and Jonas Schneider and John Schulman and Jie Tang and Wojciech Zaremba},
  title  = {OpenAI Gym},
  year   = {2016},
  eprint = {arXiv:1606.01540},
  howpublished = {arXiv},
  url    = {https://arxiv.org/abs/1606.01540}
}

@inproceedings{adam_validation,
  author    = {Adam Keane and Nick Pepper and Chris Burr and Amy Hodgkin and John Korna and Marc Thomas},
  title     = {A framework for assuring the accuracy and fidelity of an AI-enabled digital twin of en route UK airspace},
  booktitle = {AIAA SCITECH 2026 Forum},
  chapter   = {},
  pages     = {},
  doi       = {},
  url       = {},
  eprint    = {},
  year      = {2026}
}

@inproceedings{amy_aiaa,
  author    = {Amy Hodgkin and Nick Pepper and Marc Thomas},
  title     = {Conditioning Aircraft Trajectory Prediction on Meteorological Data with a Physics-Informed Machine Learning Approach},
  booktitle = {AIAA SCITECH 2026 Forum},
  chapter   = {},
  pages     = {},
  doi       = {},
  url       = {},
  eprint    = {},
  year      = {2026}
}

@inproceedings{RL_paper,
  title     = {Online Action-Stacking Improves Reinforcement Learning Performance for Air Traffic Control},
  author    = {Ben Carvell and George {De Ath} and Eseoghene Benjamin and Richard Everson},
  booktitle = {AIAA SCITECH 2026 Forum},
  chapter   = {},
  pages     = {},
  doi       = {},
  url       = {},
  eprint    = {},
  year      = {2026}
}

@inproceedings{Agent_validation_framework,
  author    = {Ben Carvell and Marc Thomas and George {De Ath} and Richard Everson and Nick Pepper and Adam Keane and Richard Cannon},
  title     = {Human-In-The-Loop Testing of AI Agents for Air Traffic Control with a Regulated Assessment Framework},
  booktitle = {AIAA SCITECH 2026 Forum},
  chapter   = {},
  pages     = {},
  doi       = {},
  url       = {},
  eprint    = {},
  year      = {2026}
}

@inproceedings{mallard,
  author    = {Paul Kent and George {De Ath} and Ben Carvell and Richard Everson and Martin Layton and Allen Hart},
  title     = {A future capabilities agent for tactical Air Traffic Control},
  booktitle = {AIAA SCITECH 2026 Forum},
  chapter   = {},
  pages     = {},
  doi       = {},
  url       = {},
  eprint    = {},
  year      = {2026}
}

@inproceedings{hawk,
  author    = {Elhassan Mohamed and Ben Carvell and Rob Procter and George {De Ath} and Richard Everson},
  title     = {Towards Transparent AI for Air Traffic Control},
  booktitle = {AIAA SCITECH 2026 Forum},
  chapter   = {},
  pages     = {},
  doi       = {},
  url       = {},
  eprint    = {},
  year      = {2026}
}

@inproceedings{Assimilation_paper,
  author    = {Nick Pepper and Marc Thomas and Zack Xuereb Conti},
  title     = {Fast Surrogate Models for Adaptive Aircraft Trajectory Prediction in En route Airspace},
  booktitle = {AIAA SCITECH 2026 Forum},
  chapter   = {},
  pages     = {},
  doi       = {},
  url       = {},
  eprint    = {},
  year      = {2026}
}

\end{document}